\documentclass[12pt,a4paper,oneside]{article}
\usepackage{graphicx,amssymb,amsmath,slashed,cite}
\usepackage[margin=2cm]{geometry}
\usepackage{array}

\newcommand{\beq}{\begin{eqnarray}}
\newcommand{\eeq}{\end{eqnarray}}

\newcommand{\lag}{\mathcal{L}}

\newcommand{\re}{\mathrm{Re}}

\newcommand{\eps}{{\epsilon}}
\newcommand{\hc}{{\rm h.c.}}

\newcommand{\footnoteremember}[2]{
\footnote{#2}
\newcounter{#1}
\setcounter{#1}{\value{footnote}}
}
\newcommand{\footnoterecall}[1]{
\footnotemark[\value{#1}]
}

\begin{document}

\title{ {\tiny \vspace*{-2cm} \hspace*{14cm} CERN-PH-TH/2013-045} \vspace*{.9cm}\\ \bf \Large Modified Higgs Physics from Composite Light Flavors}

\author{\fontsize{12}{16}\selectfont C\'edric Delaunay$^1$, Christophe Grojean$^{1, 2}$ and Gilad Perez$^{1, 3}$ \vspace{6pt}\\
\fontsize{11}{16}\selectfont\textit{$^1$ CERN Physics Department, Theory Division,
CH-1211 Geneva 23, Switzerland}\\
\fontsize{11}{16}\selectfont\textit{$^2$ IFAE, Universitat Aut\`onoma de Barcelona, 08193 Bellaterra, Barcelona, Spain}\\
\fontsize{11}{16}\selectfont\textit{$^3$ Department of Particle Physics and Astrophysics, Weizmann Institute of Science,}\\
\fontsize{11}{16}\selectfont\textit{Rehovot 76100, Israel}}
\date{}
\maketitle

\begin{abstract}
We point out that Higgs rates into gauge bosons can be significantly modified in composite pseudo Nambu--Goldstone boson (pNGB) Higgs models if quarks belonging to the first two generation are relatively composite objects as well. 
Although the lightness of the latter {\it a priori} screen them from the electroweak symmetry breaking sector, we show, in an effective two-site description, that their partners can lead to order one shifts in radiative Higgs couplings to gluons and photons.  Moreover, due to the pseudo-Goldstone nature of the Higgs boson, the size of these corrections is completely controlled by the degree of compositeness of the individual light quarks. The current measurements of flavor-blind Higgs decay rates at the LHC thus provide an indirect probe of the flavor structure of the framework of pNGB Higgs compositeness. 
\end{abstract}

\section{Introduction}

Nature seems to unitarize longitudinal electroweak (EW) boson scattering with a Higgs boson of mass around $125\,$GeV~\cite{July4,Higgs_disco,Higgs2013}. The (more than ever) burning theoretical question remains to understand why this light scalar is light. A plausible explanation is that the Higgs field is a bound state of a new dynamics which becomes strongly coupled about the TeV scale~\cite{CHM}. The composite Higgs has a characteristic size set by the strong dynamics scale so that its mass is insensitive to unknown physics at very short distances. In order to account for the little hierarchy between the observed Higgs mass and its size, an appealing possibility is to assume that the composite Higgs field is
a Nambu--Goldstone boson (NGB) of a global symmetry of the strong dynamics~\cite{HPGB, MCHM, SILH}. 
A Higgs potential is then radiatively generated through the mechanism of partial compositeness~\cite{PC}. The theory consists of linear mass mixings between the strong dynamics and the elementary Standard Model (SM) degrees of freedom, which explicitly break the global symmetry. As a result massive SM states such as the top quark and the electroweak gauge bosons are both the source and the beneficiary of EW symmetry breaking (EWSB).
Being the most massive of all SM fields the top quark typically controls the Higgs potential and drives EWSB. In order to ensure that the Higgs mass divergence induced by the top are softened by the strong dynamics, which in turn guarantees the naturalness of the EW scale, the composite partners of the top quark, the so-called top partners, must be relatively light, typically below the TeV scale~\cite{lighttoppartner,topfriends,PomRiva}. 

In contrast to the top sector, the first two quark generations (and leptons) are {\it a priori} too light to play any significant role in EWSB. Consequently,  naturalness considerations do not constrain the spectrum of their composite partners, which could be around the TeV scale as well or anywhere above it. 
Naturalness considerations do not also shed light on whether the breaking of the flavor symmetries of the strong dynamics, as well as the degree of compositeness of the light quark flavors, are large or small.  The strong dynamics can thus display a variety of flavor structures ranging from completely anarchic (all flavor symmetries are badly broken~\cite{NeuGross,GhergPom,Huber}) through U(1)$^3$~\cite{CsakiWeiler,5DMFV,shining}, approximate U(2) ~\cite{MFV,FTRS,Barbieri:U2} and up to U(3) symmetric flavor parameters~\cite{RediWeiler}. 
Light quark flavors are elementary in the anarchic approach which provides an explanation for the SM flavor hierarchies together with a mechanism to suppress new contributions to flavor and CP violating processes~\cite{APS}. However, in the presence of custodial symmetry~\cite{Zbb}, weak isospin singlet light quarks are allowed to be relatively composite in flavor symmetric models without conflicting with EW precision measurements at LEP~\cite{FTRS,RediWeiler}. Although dijets searches at the LHC~\cite{LHCdijets} already put some constraints on the compositeness of the up and down quarks~\cite{Pomarol-dijets}, second generation quarks are basically unconstrained~\cite{Exhil}.
Furthermore, the compositeness of (some of) the light quarks could be motivated by recent anomalies in the up sector data. For instance, the anomalously large forward-backward asymmetry observed in top quark pair production at the Tevatron~\cite{AFBexp,LeptonAFBexp} could point toward a relatively composite right-handed (RH) up quark in the composite Higgs  framework~\cite{FTRS2}. Similarly the surprisingly large direct CP asymmetry in singly Cabibbo suppressed $D$ meson decays~\cite{HFAG}, first observed at LHCb~\cite{DACPexpLHCb} and further supported by other experiments~\cite{DACPexpOthers}, can be accommodated in composite Higgs models where the RH up, charm and strange quarks are composite as well~\cite{Exhil}.
However, at present it is hard to draw a definitive conclusions weather the above measurements are to be interpreted as a sign for non-SM dynamics~(see~{\it e.g.}~Refs.~\cite{BKZupan, MFMWPerez}). Furthermore, due to small sea quark luminosities, it is rather challenging quite generically to probe for second generation compositeness at the LHC era using direct searches.

We identify in this paper a new type of Higgs couplings modification in the composite NGB Higgs framework which arises from light quark flavor compositeness. We also show that under reasonable assumptions Higgs rates at the LHC could significantly deviate from their SM expectations in the presence of composite light quark flavors.
In particular sizable effects can arise in flavor blind observables such as Higgs production cross-sections through gluon fusion and Higgs branching ratios into diphotons and weak bosons. Thus, flavor conserving Higgs data could provide,  rather surprisingly, a unique window on composite flavor physics and possibly probe, at least at the qualitative level, the flavor structure of the strong dynamics. 
Furthermore, we show that Higgs coupling corrections arising from composite light quark flavors are comparable in size to the well-known corrections from a composite Higgs~\cite{SILH} and that the former can be used to hide the composite nature of the Higgs boson in single Higgs production at the LHC.\\

The remainder of the paper is organized as follows. In Section~\ref{sec:Hcpls}, we derive the effect of composite light quarks on radiative Higgs couplings to gluons and photons in a broad class of models where the SM quarks mix with heavy vector-like quarks that couple directly to the Higgs. Then we focus on the case where the Higgs particle is a pseudo NGB (pNGB).
Section~\ref{sec:flavor} reviews the possible structures of the strong sector flavor parameters. We discuss phenomenological implications of composite flavors on Higgs rates at the LHC in Section~\ref{sec:pheno} and present our conclusions in Section~\ref{sec:conclu}.

\section{Modified Higgs Couplings from Composite Flavors} \label{sec:Hcpls}

\begin{figure}[t]
\centering
\begin{tabular}{ccc}
\includegraphics[scale=0.65]{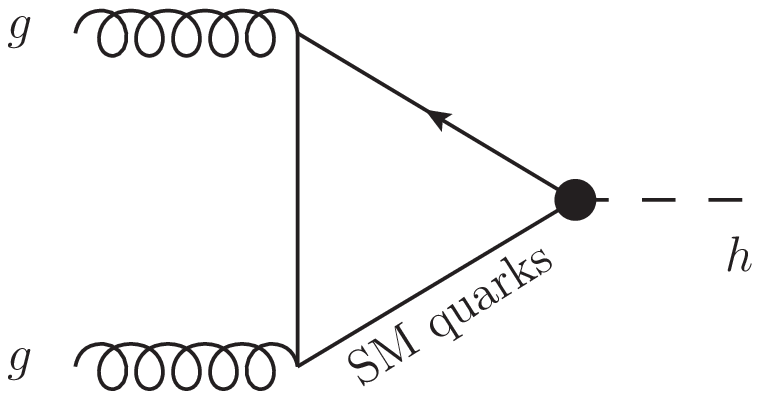}&\quad\quad\quad&
\includegraphics[scale=0.65]{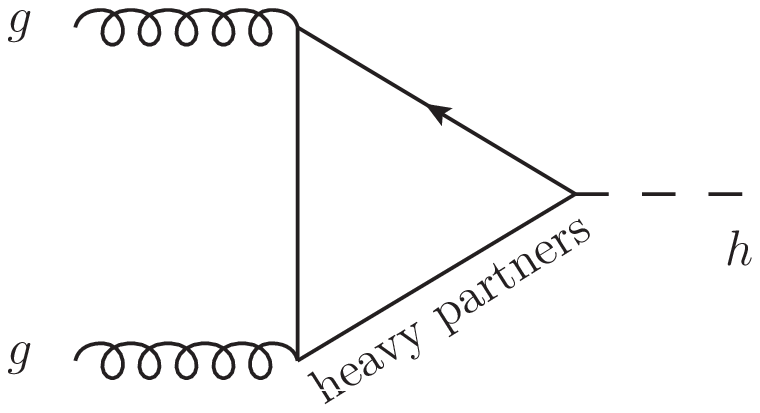}\\
 (a) &   & (b) 
\end{tabular}
\caption{Generic one-loop diagrams contributing to the gluon fusion Higgs production NP amplitude in the presence of composite fermionic resonances. Mass eigenstates are understood in the loops. Diagram (a) is the SM quark contribution where the black filled circle denotes the modified Yukawa coupling, whose deviation from the SM value is caused both by mixing with the composite states and possibly by Higgs non-linearities. Diagram (b) is the contribution from heavy resonances. 
}
\label{diag_hgg}
\end{figure}

We study here the effects of  vector-like fermionic partners of the SM quarks on Higgs couplings in models of partial compositeness. Part of the results presented below were already pointed out in the literature but we find it useful to review them in a more general context where light SM flavors also have large couplings to their partners. Fermionic resonance contributions to the Higgs couplings to photons and gluons are two fold. There is a direct one-loop contribution where the resonances themselves run in the loop (see Fig.~\ref{diag_hgg}b), and an indirect contribution arising from a modification of the SM Yukawa couplings in the SM loops due to mixing with the resonances~\cite{AdamHgg, LowRattazzi, LowVichi, AzatovGal,Gillioz:2012se} (see Fig.~\ref{diag_hgg}a).
%

We will use an effective field theory (EFT) approach below the resonances mass scale in order to describe the two effects. Then, we will discuss generic results in a simple two-site model and finally move to MCHM where the Higgs is realized as a pNGB.

\subsection{EFT below the resonances}

We rely on the following effective Lagrangian to describe the Higgs coupling to SM fermions and gauge bosons below the composite resonance mass scale~\cite{SILH}
\beq\label{Leff}
\lag_{\rm eff} & = \lag_{\rm SM} + \lag_{\rm NP}\,,
\eeq
with
\beq
\lag_{\rm SM} &\supset&  i\bar q_L\slashed D q_L + i\bar u_R\slashed D u_R +i\bar d_R\slashed D d_R - \left(y_u \bar q_L \tilde H u_R +y_d \bar q_L H d_R+ \hc\right) \nonumber
\eeq
and
\beq
\lag_{\rm NP}=  \sum_i c_i\mathcal{O}_i &\supset& c_r|H|^2 |D_\mu H|^2+\frac{c_H}{2}\partial_\mu(H^\dagger H)\partial^\mu(H^\dagger H)+ \frac{\alpha_s c_g}{12\pi} |H|^2 G_{\mu\nu}^{a\,2}+ \frac{\alpha c_\gamma}{2\pi} |H|^2 F_{\mu\nu}^{2}\nonumber \\
&& +y_u c_{y_u} \bar q_L \tilde H u_R |H|^2+y_d c_{y_d} \bar q_L  H d_R |H|^2+\hc\,,\label{LNP}
\eeq
where $D_\mu$ is the SM covariant derivative, $H$ is the SM Higgs doublet and $\tilde H=i\sigma_2 H^*$, $F_{\mu\nu}$ and $G_{\mu\nu}^a$ are the photon and gluon field strength tensors ($\alpha$ and $\alpha_s$ are the  QED and QCD coupling strengths) and $q_L$ and $u_R$, $d_R$ are the SU(2)$_L$ quark doublet and up- and down-type singlets. Flavor indices are implicit. $\lag_{\rm SM}$ is the SM Lagrangian and we only consider a subset of mass dimension six operators in $\lag_{\rm NP}$ which are relevant to the analysis performed in the remainder of the paper.

The operators $\mathcal{O}_{r}$ and $\mathcal{O}_{H}$ in Eq.~\eqref{LNP} are required to capture non-linear Higgs effects in models where the Higgs field is realized as a pNGB. These two operators are redundant and do not yield independent on physical observables~\cite{BuchWyler,Polishgroup}. However, we keep both present since this provides us with a convenient operator basis for MCHMs.\footnoteremember{cr}{One can move, by redefining the Higgs field, to an operator basis where $c_r=0$ (the so-called SILH basis~\cite{SILH}), while other coefficients shift as $c_H\to c_H-c_r$ and $c_{y_{u,d}}\to c_{y_{u,d}}+c_r/2$, and $c_{g,\gamma}$ remain unchanged.}  $\mathcal{O}_{y_{u,d}}$ parameterize the modifications of the SM Yukawa couplings, which receive contributions both from Higgs non-linearities and the presence of vector-like fermions, while  $\mathcal{O}_{g,\gamma}$  are only induced by the latter. We rescaled the coefficients of $\mathcal{O}_g$ and $\mathcal{O}_\gamma$ to account for the fact that these operators are induced at least at one-loop. Furthermore we normalized their coefficients so that for $c_g=1/v_{\rm SM}^2$ and $c_\gamma=Q_u^2/v_{\rm SM}^2$, where $Q_u=2/3$ is the up-type quark electric charge and $v_{\rm SM}=(\sqrt{2}G_F)^{-1/2}\simeq 246$~GeV, the $\mathcal{O}_g$ and $\mathcal{O}_\gamma$ respective contributions to the dimension five operators $hG_{\mu\nu}^{a\,2}$ and $hF_{\mu\nu}^{2}$ at the Higgs mass scale, where $h$ is the physical Higgs boson, are of the same magnitude as the SM top contributions.

We assumed for simplicity that NP is CP conserving so that CP odd operators like $|H|^2G_{\mu\nu}^a\widetilde G^{\mu\nu\,a}$ or $|H|^2F_{\mu\nu}\widetilde F^{\mu\nu}$, with $\widetilde G$ and $\widetilde F$ the dual gauge field strength tensors,  are not induced and $c_{y_{u,d}}$ are real. Also, we did not write explicitly dimension six operators like $\mathcal{O}_{\slashed Du}=i(\bar u_R \slashed D u_R) |H|^2$, $\mathcal{O}_{\slashed Dd}=i(\bar d_R \slashed D d_R) |H|^2$ and $\mathcal{O}_{\slashed Dq}=i(\bar q_L \slashed D q_L) |H|^2$ as they are also redundant.\footnoteremember{redef}{
It is always possible to reach an operator basis where $c_{\slashed Du}=c_{\slashed Dd}=c_{\slashed Dq}=0$ by mean of quark field redefinitions under which only $c_{y_{u,d}}$ shift as $c_{y_{u,d}}\to c_{y_{u,d}} +c_{\slashed Du,\slashed Dd}^*+c_{\slashed Dq}$.} 
We omitted the custodial symmetry breaking operators $\mathcal{O}_T=|H^\dagger D_\mu H|^2$, $\mathcal{O}_{u}=\bar u_R \gamma^\mu u_R (H^\dagger D_\mu H)$, $\mathcal{O}_{d}=\bar d_R \gamma^\mu d_R (H^\dagger D_\mu H)$, $\mathcal{O}_{ud}=\bar u_R \gamma^\mu d_R( \tilde H^\dagger D_\mu H)$, $\mathcal{O}_{q}^1=\bar q_L \gamma^\mu q_L (H^\dagger D_\mu H)$ and $\mathcal{O}_{q}^3=\bar q_L \sigma^a \gamma^\mu q_L (H^\dagger \sigma^aD_\mu H)$. These operators cannot be removed by field redefinitions and yield independent physical effects. In particular they modify the $\rho$ parameter and the SM quark couplings to the $Z$ boson, which were all precisely measured at LEP up to the per mile accuracy, see {\it e.g.} Ref.~\cite{Gfitter}. As already mentioned, we focus below only on models where the strong dynamics is SO(4) invariant and where the right-handed (RH) and left-handed (LH) elementary quarks mix with composite fermions in the SO(4) singlet and fundamental representations, respectively. In such a case $\mathcal{O}_T$, $\mathcal{O}_{u}$, $\mathcal{O}_{d}$ and $\mathcal{O}_{ud}$ are not induced. The custodial symmetry does not however prevent $\mathcal{O}_{q}^1$ and $\mathcal{O}_{q}^3$  to arise, but only guarantees that the net shift to the $Z$ coupling of one weak isospin component of $q_L$ vanishes, leaving the other component unprotected. 
However, these operators can only arise through mixing with the composite sector, therefore of crucial importance only when dealing with the LH bottoms. 
But, as long as the left handed (LH) light quarks are mostly elementary, as assumed below, these operators can be safely neglected.
Finally, we do not consider dipole-like operators such as $\bar q_L H\sigma^{\mu\nu} T^a u_R G_{\mu\nu}^a$, which contribute to radiative Higgs couplings at the one-loop level. These operators are expected to arise at loop-level in MCHM~\cite{SILH}, so their effects are typically subdominant and we neglect them here (see {\it e.g.} Refs.~\cite{KPW,contino} for a dedicated discussion).\\

EWSB is switched on by taking (in unitary gauge) $H^T\to (0,(v+\hat h)/\sqrt{2})$ where $v$ is the Higgs vacuum expectation value (VEV) and $\hat h$ is a neutral parity even fluctuation. Note that $v$ is related to the Fermi constant $G_F$ through
\beq
v= v_{\rm SM}\left(1-c_r\frac{v_{\rm SM}^2}{4}\right)+\mathcal{O}(v_{\rm SM}^5)\,,\quad v_{\rm SM} \equiv \left(\sqrt{2}G_F\right)^{-1/2} \simeq 246\,{\rm GeV}\,,
\eeq
and $\hat h$ is not canonically normalized. The physical Higgs boson $h$ with a canonical kinetic term is 
\beq
h= \left[1+\left(c_H+\frac{c_r}{2}\right)\frac{v^2}{2}+\mathcal{O}(v^4)\right]\hat h + \mathcal{O}(\hat h^2)\,.
\eeq

The above effective Lagrangian yields the following linear interaction of the Higgs boson with fermion bilinears
\beq
-\lag_{h\bar f f } = \frac{m_u}{v_{\rm SM}}\left[1-\left(\re [c_{y_u}]+\frac{c_H}{2}\right)v^2+\mathcal{O}(v^4)\right]h\bar u u+\{u\to d\} \,,
\eeq
where 
\beq
\label{masses}
m_{u,d}=y_{u,d}\frac{v}{\sqrt{2}}\left(1-c_{y_{u,d}}\frac{v^2}{2}\right) 
\eeq
are the SM-quark masses. $\lag_{\rm NP}$ 
contributes to the gluon fusion Higgs production amplitude as~\cite{GeorgiHgg}
\beq
\mathcal{M}_{gg\to h} \propto  c_g v^2+\sum_{i=u,d}\left[1- \left({\rm Re}[c_{y_i}]+\frac{c_H}{2}\right)v^2\right]A_{1/2}(\tau_i)\,,
\label{Mhgg}
\eeq
where $A_{1/2}$ is a fermion loop function (see Appendix~\ref{Loops}) which only depends on the SM-quark mass $m_{u,d}$ 
and the Higgs boson mass $m_h$ through $\tau_{u,d}\equiv m_h^2/(4m_{u,d}^2)$. For a heavy  flavor like the top quark, one has $m_u\gg m_h/2$ and the loop function asymptotes to $A_{1/2}(0)= 1$ so that the top partners contribution is just $\propto  c_g-{\rm Re}[c_{y_u}]$. On the other hand, for a light SM flavor with $m_u\ll m_h/2$, one has $A_{1/2}(\infty) = 0$ and the associated resonances only affects Higgs production through $ c_g$. 

The Lagrangian $\lag_{\rm NP}$ also corrects the Higgs to two photons decay amplitude~\cite{EllisGaillard,LET} 
\beq\label{Mhyy}
\mathcal{M}_{h\to \gamma\gamma} \propto  c_\gamma v^2- \frac{7}{4}\left[1+\left(c_r-c_H \right)\frac{v^2}{2}\right]A_1(\tau_W)+\sum_{i=u,d}Q_i^2\left[1- \left({\rm Re}[c_{y_i}]+\frac{c_H}{2}\right)v^2\right]A_{1/2}(\tau_i)\,,
\eeq
where $Q_u=2/3$ and $Q_d=-1/3$ are the up- and down-type quark electric charges, $A_1$ is the $W$ loop function (see Appendix~\ref{Loops}) and $\tau_W\equiv m_h^2/(4m_W^2)$. Finally, the tree-level induced Higgs to $ZZ^*$, $WW^*$ and $u\bar u$, $d\bar d$ decay amplitudes are modified as
\beq\label{Mhothers}
\mathcal{M}_{h\to ZZ,WW} \propto 1+\left(c_r-c_H \right)\frac{v^2}{2}\,,\quad \mathcal{M}_{h\to u\bar u,d\bar d} \propto 1-\left(2{\rm Re}[c_{y_{u,d}}]+c_H\right)\frac{v^2}{2}\,.
\eeq

We  match in the following the effective Lagrangian Eq.~\eqref{Leff} to  NP models with vector-like fermions. We begin by studying a toy model where SM chiral quarks mix with vector-like fermions with the same SM quantum numbers, and then move to the more realistic MCHM. 

\subsection{A Two-Site Toy Model}

\begin{figure}[t]
\centering
\includegraphics[scale=0.36]{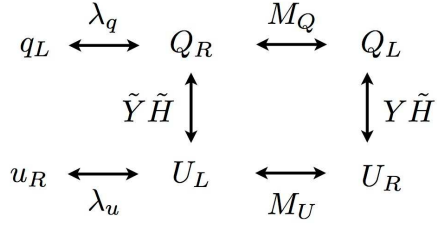}
\caption{Two-site model: the elementary quarks, $q_L$ and $u_R$, mix with vector-like massive quarks, $Q$ and $U$, that belong to the composite sector and have Yukawa interaction with the Higgs field.}
\label{fig:moose}
\end{figure}

As a warm-up we consider a simple toy model where the Higgs field only has linear couplings to fermions. For simplicity, we focus on a single up-type flavor and we add one vector-like SU(2)$_L$ doublet $Q$ and one vector-like singlet $U$ to the SM chiral quark doublet $q_L$ and singlet $u_R$. The relevant Lagrangian is (see fig.~\ref{fig:moose})
\beq\label{L2site}
\lag_{{\rm toy}} &=&  i\bar q_L\slashed D q_L + i\bar u_R\slashed D u_R + \bar Q\left(i\slashed D-M_Q\right)Q+\bar U \left(i\slashed D-M_U\right)U\,,\nonumber\\
&&-Y\bar Q_L \tilde HU_R-\tilde Y\bar Q_R \tilde H U_L  -\lambda_q \bar q_L Q_R -\lambda_u \bar U_L u_R+\hc
\eeq
Following the partial compositeness approach, we assume that chiral fermions do not directly couple to the Higgs doublet $H$. Rather, EWSB is mediated to chiral fields through their linear mass mixing, $\lambda_{q,u}$, to vector-like fermions.

\begin{figure}[!p]
\centering
\begin{tabular}{c c c m{0.3cm} c}
\raisebox{-.5\height}{\includegraphics[scale=0.58]{Dlight.ps}} &  $=$ & \raisebox{-.5\height}{\includegraphics[scale=0.58]{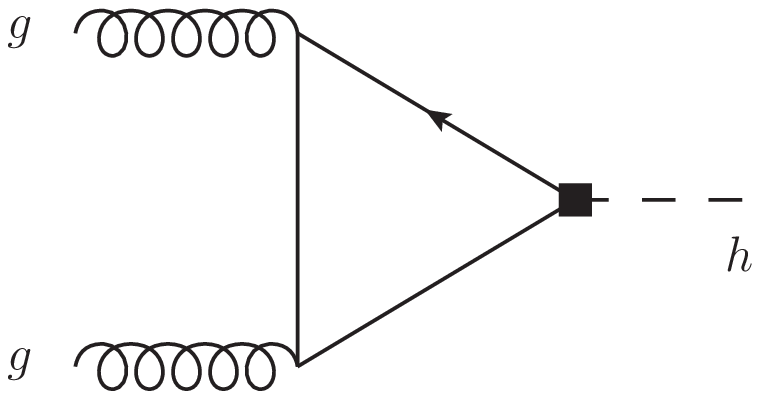}}  & $+$ & \raisebox{-.5\height}{\includegraphics[scale=0.58]{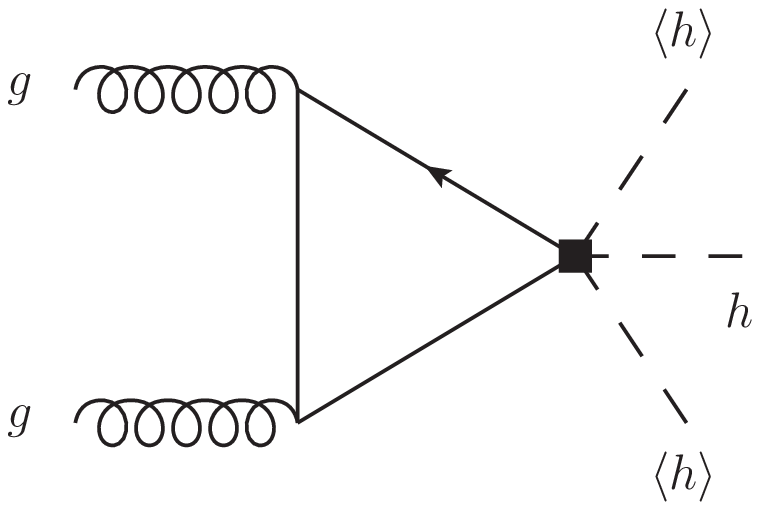}}  \\
(a) & & (a1) & & (a2) \\
& $+$ & \raisebox{-.5\height}{\includegraphics[scale=0.58]{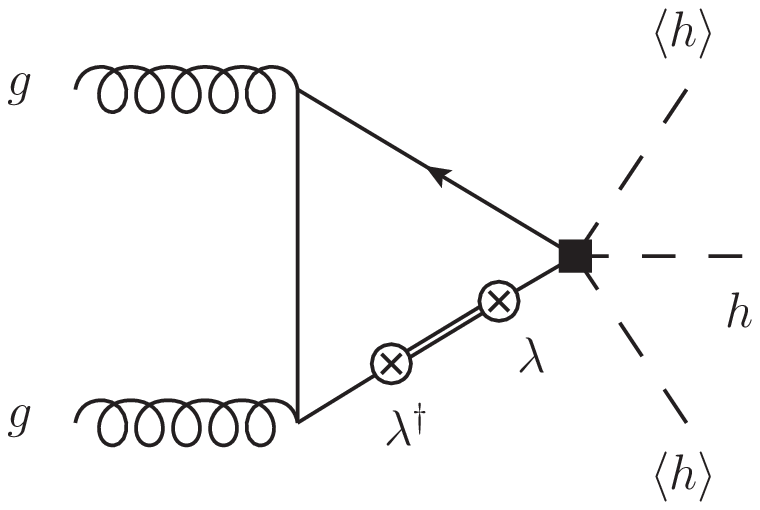}} & $+\quad$ & $ \cdots$ \\
  &  & (a3) & & \\
\\
\\
\raisebox{-.5\height}{\includegraphics[scale=0.58]{Dheavy.ps}} &  $=$ & \raisebox{-.5\height}{\includegraphics[scale=0.58]{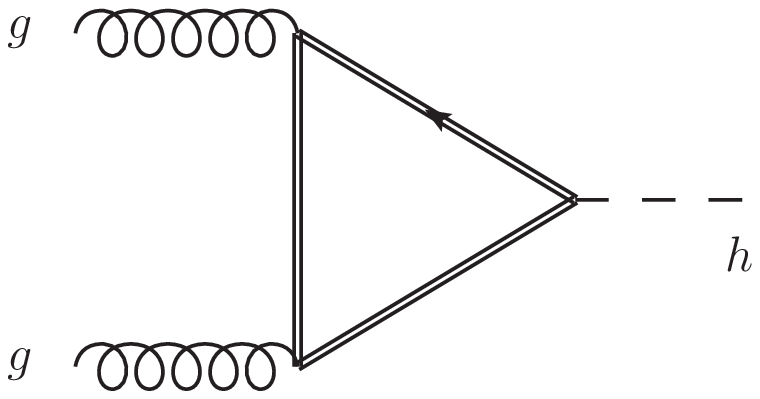}}  & $+$ & \raisebox{-.5\height}{\includegraphics[scale=0.58]{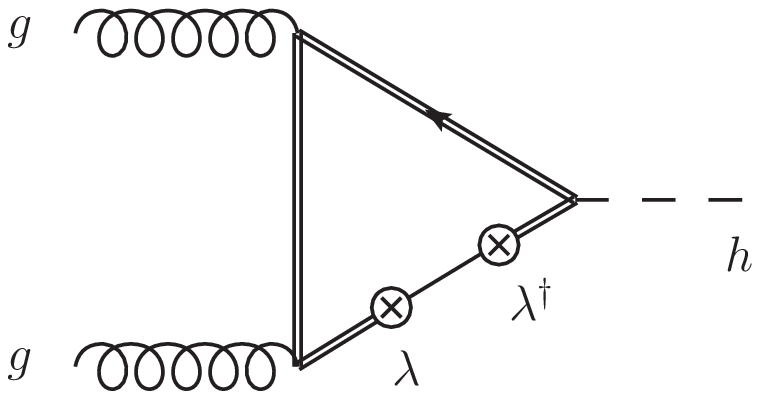}}  \\
(b) & & (b1) & & (b2)\\
\\
& $+$ & $\cdots$ & & 
\end{tabular}
\caption{Leading contributions to the generic diagrams of Fig.~\ref{diag_hgg}, in an expansion in small elementary/composite mixings, $\lambda_x$ with $x=q^u,q^d,u,d$. We use in the text expressions valid to all orders. Non-linear Higgs interactions arising in pNGB models are not represented. Mass eigenstates are understood on the left-hand side of the equalities, whereas, on the right-hand side, single and double lines stand for elementary quarks and composite resonances, respectively, and the crossed-circle denotes $\lambda_x$ insertions. Black squares are effective Yukawa interactions for the elementary fields generated  through mixings (see Fig.~\ref{diag_tree}). Dots denote diagrams of higher order in $\lambda_x$. 
Diagram (a1) is the SM contribution, while diagrams (a2) and (a3) are corrections due to mixing with the composite resonances. Diagrams (b1) and (b2) are contributions from the composite resonances. For the top sector, diagrams (a2) and (a3) cancels, to leading order in $m_h^2/(4m_t^2)$, against diagrams (b1) and (b2), respectively, provided the determinant of the associated mass matrix can be factorized as in Eq.~\eqref{detMfacto}. For light quark generations, diagram (a) is $4m_q^2/m_h^2$ suppressed and the composite partners yield large contributions through diagrams (b1) and (b2). 
In pNGB Higgs model, diagrams (a2) and (b1) vanish individually for all flavors due to the global symmetry of the Goldstone bosons.}
\label{diagrams}
\end{figure}

\begin{figure}[t]
\centering

\begin{tabular}{ccccccc}
\raisebox{-.22\height}{\includegraphics[scale=0.5]{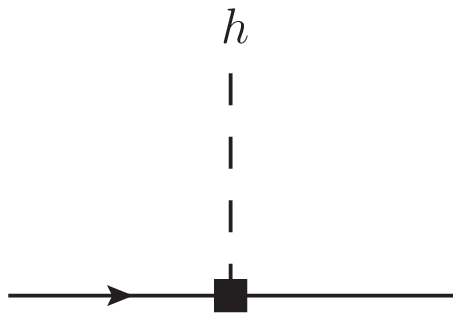}} &  $=$ & \raisebox{-.26\height}{\includegraphics[scale=0.5]{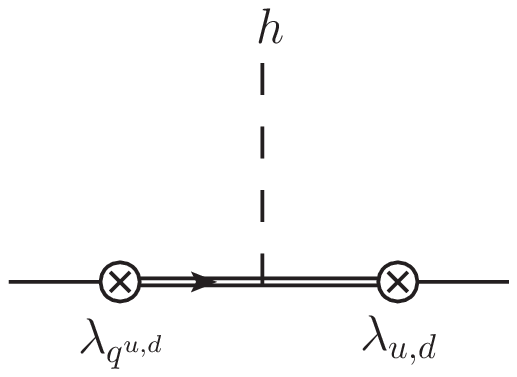}}&\,,&
\raisebox{-.22\height}{\includegraphics[scale=0.5]{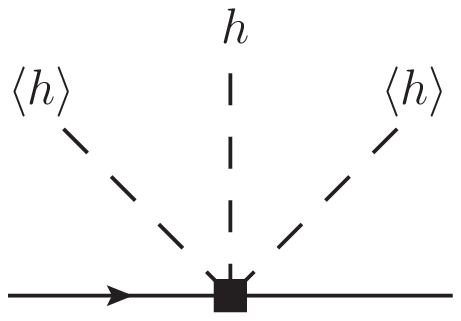}} &  $=$ & \raisebox{-.26\height}{\includegraphics[scale=0.5]{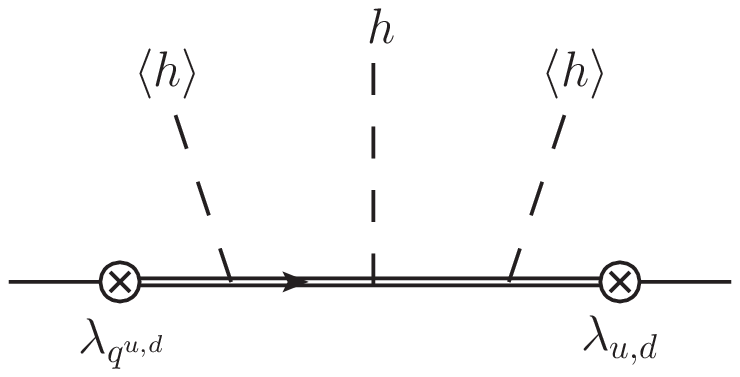}}
\end{tabular}
\caption{Tree-level diagrams generating the effective vertices used in Fig.~\ref{diagrams}. Single and double lines stand for elementary quarks and composite resonances, respectively, and the crossed-circle denotes elementary/composite mixing insertion.
Non-linear Higgs interactions arising in pNGB models are not represented.}
\label{diag_tree}
\end{figure}

We now match $\lag_{\rm toy}$ onto the effective Lagrangian in Eq.~\eqref{Leff}. At tree-level we find $c_r=c_H=0$, since the Higgs is linearly realized, while integrating out the vector-like fermions  yields~\cite{BurasVF}
\beq
y_u= Y\sin\theta_q\sin\theta_u\,,
\eeq
and
\beq\label{cyutoy}
c_{y_u}=-\frac{Y\tilde Y^*}{\tilde M_Q \tilde M_U}\cos\theta_q\cos\theta_u +\frac{|Y|^2}{2\tilde M_Q^2}\cos^2\theta_q\sin^2\theta_u+\frac{|Y|^2}{2 \tilde M_U^2}\cos^2\theta_u\sin^2\theta_q\,,
\eeq
where we introduced the eigenmasses prior to EWSB $\tilde M_Q^2\equiv M_Q^2+\lambda_q^2$, $\tilde M_U^2\equiv M_U^2+\lambda_u^2$, as well as the sine and cosine of the elementary/composite mixing angles: $\sin\theta_{q,u}\equiv \lambda_{q,u}/\tilde M_{Q,U}$ and $\cos\theta_{q,u}\equiv M_{Q,U}/\tilde M_{Q,U}$. In the limit of small mixing, {\it i.e.} $\epsilon_{q,u}\equiv \lambda_{q,u}/M_{Q,U}\ll 1$, $\cos\theta_{q,u}\simeq 1+\mathcal{O}(\epsilon^2)$ and $\sin\theta_{q,u}\simeq \epsilon_{q,u}$.
Note that the last two terms of $c_{y_u}$ arise from higher dimensional quark kinetic operators $\mathcal{O}_{\slashed D u}$ and $\mathcal{O}_{\slashed D q}$, respectively, which are reshuffled in terms of $\mathcal{O}_{y_u}$ by mean of field redefinitions (see footnote\footnoterecall{redef}).
Matching the Higgs to two gluons and photons amplitudes at one-loop determines the two remaining Wilson coefficients in $\lag_{\rm NP}$~\footnote{Note that since  $\mathcal{O}_{g,\gamma}$ are CP-even operators, they are only sensitive to the real part of $Y\tilde Y^*$. The imaginary part of $Y\tilde Y^*$ would only match to their CP-odd counterparts.}
\beq\label{cg2site}
 c_g = Q_u^{-2}c_\gamma = -\frac{{\rm Re}[Y\tilde Y^*]}{\tilde M_Q\tilde M_U}\cos\theta_q\cos\theta_u+\frac{|Y|^2}{2 \tilde M_Q^2}\cos^2\theta_q\sin^2\theta_u+\frac{|Y|^2}{2\tilde M_U^2}\cos^2\theta_u\sin^2\theta_q\,.
\eeq
In the limit of small  mixings, the contribution of the top and its partners depicted in Fig.~\ref{diag_hgg} can be diagrammatically understood from the expansion detailed in Figs.~\ref{diagrams} and \ref{diag_tree}.


Several comments are in order: 
\begin{itemize}

\item We find the following relations to hold: $c_g=\re [c_{y_u}]$ and $c_\gamma =Q_u^{2}\re [c_{y_u}]$. 
Examining Eqs.~\eqref{Mhgg} and~\eqref{Mhyy} we find that there are no net effects on radiative Higgs couplings from the top partners. This cancelation, which was already observed in pNGB Higgs models~\cite{AdamHgg,AzatovGal}, is not related to pNGB symmetries. It is straightforward (see {\it e.g.} Ref.~\cite{Gillioz:2012se}) to use the low-energy Higgs theorems (LEHT)~\cite{LET,spira} to 
formulate a general condition for a model to enjoy this cancelation.
For models involving heavy fermions, $m_f\gg m_h/2$, the contribution of the latter to Higgs radiative couplings is $\propto \partial_{\,\log v}\log\det\mathcal{M}$, where $\mathcal{M}$ is the fermion mass matrix (see {\it e.g.} Ref.~\cite{AzatovGal}). Therefore, as long as the determinant of the mass matrix can be factorized as 
\beq\label{detMfacto}
\det \mathcal{M}= F(v/f)\times P(Y, M,f)\,,
\eeq
where $F(0)=0$, $f$ is the Higgs decay constant of pNGB models, and $Y$ and $M$ stand for the heavy fermion Yukawa couplings and masses respectively, Higgs rate to gluons and photons would not get any correction from the presence of the heavy top partners.
Moreover, in the special case where $F(v/f)\propto v$ the models's predictions coincide with that of the SM. The model defined in Eq.~\eqref{L2site} falls in this class, since the quark mass matrix~\footnote{The fact the light quark mass in Eq.~\eqref{masses}, which depends on the matched value of $c_{y_u}$, must be an eigenvalue of $\mathcal M$ provides an independent check of Eq.~\eqref{cyutoy}.} 
\beq
\mathcal{M}=\left(\begin{array}{ccc} 
0 & \lambda_q & 0 \\ 0 & M_Q & Y v/\sqrt{2} \\ \lambda_u & \tilde Y v/\sqrt{2} & M_U
\end{array}\right)\,
\eeq
has a determinant which is only linear in the Higgs VEV, $\det\mathcal{M}=Y\lambda_q\lambda_u v/\sqrt{2}$. 
 Higgs couplings to gluons and photons are thus independent of the top partner's spectrum and the top compositeness, and are SM-like, which results in the above relations between $c_{g,\gamma}$ and $c_{y_u}$ in the EFT. 
There are several ways to violate these relations. For instance, it is straigthforward to check that allowing for direct couplings between chiral quarks and the Higgs field would yield a non-zero top partner contribution.
Another possibility arises in pNGB Higgs models where the mass determinant factorizes but with $F(v/f)$ which is no longer linear in $v$ but is rather a trigonometric function of $v/f$. In this case Higgs radiative couplings are not SM-like, albeit being still independent of the top partner parameters. 

\item In the zero mixing limit ($\epsilon_{q,u}= 0$), the loop induced operators are controlled by the ``wrong chirality'' Yukawa coupling $\tilde Y$. This is again easy to understand from LEHT, since the determinant of the sub-block of $\mathcal{M}$ corresponding to the heavy states 
 only depends on the Higgs background through the $Y\tilde Y^*$ combination.\footnote{Similar results were already observed for dipole operators~\cite{dipoles} and Higgs FCNCs~\cite{HFCNC} in the composite Higgs framework.}

\item In non-pNGB Higgs  models where the Yukawas are all $\mathcal{O}(1)$ and anarchic, the $Y\tilde Y^*$ contribution to $c_{g,\gamma}$, which is not suppressed by partial compositeness, yields sizable $\mathcal{O}(1)$ effects on radiative Higgs couplings from composite partners of the first and second generations (and bottom) SM flavors, thus probing compositeness scales up to $\mathcal{O}(10)\,$TeV~\cite{AzatovRSHgg,NeubertRSHgg}. Moreover,  in this case, the net prediction of the model is obtained only after summing over a large tower of strong sector resonances\footnote{In the five dimensional dual description~\cite{RS,holog}, the contribution of the strong dynamics is supported by resonances whose mass is comparable with the inverse Higgs width in the bulk~\cite{AzatovRSHgg,NeubertRSHgg,dipoles}.} which are not captured by the two-site description of Eq.~\eqref{L2site}.  In contrast note that in more natural ({\it i.e.} less fine-tuned) models where the composite Higgs is a pNGB, the aforementioned effects do not arise, as we show explicitly below for MCHMs. This is so because the strong dynamics preserves the Goldstone symmetry of the Higgs field and thus cannot induce  non-derivative Higgs couplings at any order~\cite{SILH}. In this case the strong sector contribution is dominated by the lowest lying resonances and controlled by the elementary/composite mixings which breaks the Goldstone symmetry.

\item Notice that in Eq.~\eqref{cg2site} terms suppressed by the partial compositeness only involves the mixing of one chirality at a time, while SM masses are given as usual by their product. This is easily understood from the U(3)$_q\times$U(3)$_u$ flavor symmetries, under which $\epsilon_{q,u}$ are spurions transforming formally as (${\bf 3},{\bf 1}$) and (${\bf 1},{\bf \bar{3}}$), respectively. 
SM masses are bi-fundamentals $({\bf 3},{\bf \bar{3}})$ of the above flavor group, while obviously $c_{g,\gamma}$ are singlets. Therefore the smallest combinations of spurions that can contribute to those operators are of the form  $y\propto \epsilon_q \epsilon_u$ and $c_{g,\gamma}\propto 1 + \epsilon_q^\dagger\epsilon_q + \epsilon_u^\dagger\epsilon_u$. This observation has important implications. In anarchic models, both $\epsilon_{q,u}$ are small for the first two generations in order to account for the smallness of quark masses and CKM mixing angles. Therefore, no effect (besides $Y\tilde Y^*$) is expected on Higgs production and decay from composite partners of the first two generation quarks. However, if one light quark chirality is relatively composite, sizable effects on radiative Higgs couplings would arise, while the hierarchy of masses is ensured by the elementary nature of the other chirality. 
As we argue in the following sections, this opens the very interesting possibility that {\it flavor conserving} Higgs physics could in principle shed light on the flavor structure of the strong dynamics. 
\end{itemize}

\subsection{Composite pNGB Higgs Models}

We move now to consider models where the  Higgs fied is a composite  pNGB. For concreteness, we focus on the  SO(5)$/$SO(4) symmetry breaking coset, which is the minimal choice with a built-in custodial symmetry~\cite{MCHM}, but extension to larger cosets is also possible~\cite{NMCHM}. We use a simplified two-site description~\cite{twosite} of the model which consists of two distinct sectors. The first, so-called  elementary, sector is made of the SM chiral quarks (and SM gauge fields) which are taken to linearly mix with a set of vector-like fermions from the other, so-called composite, sector where a global SO(5)$\times$U(1)$_X$ symmetry is non-linearly realized. 
The breaking of SO(5)$\to$SO(4) occurs at the scale $f\lesssim\,$TeV and is parameterized by the VEV of a scalar field $\Sigma$ transforming as a fundamental of SO(5) with zero $X$ charge. The SM Higgs  doublet then emerges as a real fourplet of SO(4) ``pions'', which in turn breaks SO(4)$\to$SO(3) and thus EW symmetry.

The quantum numbers of fermionic resonances of the strong sector are arbitrary {\it a priori}. Yet, representations whose SO(4)$\sim$SU(2)$_L\times$SU(2)$_R$ decomposition is invariant under a  $P_{LR}$ parity exchanging the quantum numbers under SU(2)$_L$ and SU(2)$_R$ are phenomenologically more favored because of custodial symmetry~\cite{Zbb}. The first smallest irreducible SO(5) representations with this property are the ${\bf 5}$ (fundamental), ${\bf 10}$ (adjoint) and the ${\bf 14}$ (symmetric traceless $5\times 5$ matrices). For definiteness we henceforth focus on MCHMs where fermionic resonances transform as fundamental representations of SO(5). Although the top sector contribution can be qualitatively different and the independence of the $gg\to h$ or $h \to \gamma \gamma$ rates on the top partners spectrum is not a general feature of pNGB models, we show in Appendix~\ref{1014matching} that other choices of representation could lead to similar structure and hence result in  qualitatively similar contribution from composite light flavors.\\ 

The relevant two-site Lagrangian is 
\beq\label{LMCHM}
\lag_{2{\rm site}} = \lag_{\rm strong} + \lag_{\rm elem} + \lag_{\rm mix}\,, 
\eeq
where the elementary sector, strong sector, and elementary/composite mixing parts are respectively (flavor indices are understood)
\beq
\lag_{\rm elem}& = & i\bar q_L \slashed D q_L + i \bar u_R\slashed D u_R+i\bar d_R \slashed D d_R\,,\\
\lag_{\rm strong} &= & \frac{f^2}{2}D_\mu \Sigma(D^\mu \Sigma)^\dagger + \sum_{i=u,d}\bar{\Psi}^i\left(i\slashed D-M_i\right)\Psi^i - Y_i f\left(\bar \Psi^i_L \Sigma^T\right)\left(\Sigma\Psi^i_R\right)+\hc\,,\\
-\lag_{\rm mix} &=& \lambda_{q^u}\, \bar{q}_L D^u_{\frac{1}{6}} +\lambda_{q^d}\, \bar{q}_L D^d_{\frac{1}{6}} 
+ \lambda_u\, \bar S_{\frac{2}{3}} {u}_R + \lambda_d\, \bar S_{-\frac{1}{3}} {d}_R +\hc\,,\label{Lmix}
\eeq 
where $\lag_{\rm strong}$ is manifestly invariant under the global SO(5)$\times$U(1)$_X$ symmetry. $\Psi^u$ and $\Psi^d$ are vector-like composite fermions which we assume to live in ${\bf 5}_\frac{2}{3}$ and ${\bf 5}_{-\frac{1}{3}}$ representations of SO(5)$\times$U(1)$_X$, respectively (two composite fields $\Psi^u$ and $\Psi^d$ are necessary to generate a mass to both the up- and down-type quarks).
Recall the ${\bf 5}$ of SO(5) decomposes as ${\bf 4}+{\bf 1}$ under the unbroken SO(4) and $\Psi^{u,d}$ each consists of two SU(2)$_L$ doublets $D_\mathcal{Y}$ of hypercharge  $\mathcal{Y}=X\pm 1/2$ and an SU(2)$_L$ singlet $S_{\mathcal{Y}}$ of hypercharge $\mathcal{Y}=X$. 
These states are embedded into $\Psi^{u,d}$ as (see Appendix~\ref{Irreps})
\beq
\Psi^u& =& \frac{1}{\sqrt{2}}\left(D_\frac{1}{6}^{u\,-}-D_\frac{7}{6}^+,-i\left(D_\frac{1}{6}^{u\,-}+D_\frac{7}{6}^+\right),D_\frac{1}{6}^{u\,+}+D_\frac{7}{6}^-,i\left(D_\frac{1}{6}^{u\,+}-D_\frac{7}{6}^-\right),\sqrt{2}S_\frac{2}{3}\right)^T\,,\\
\Psi^d& =& \frac{1}{\sqrt{2}}\left(D_{-\frac{5}{6}}^--D_\frac{1}{6}^{d\,+},-i\left(D_{-\frac{5}{6}}^-+D_\frac{1}{6}^{d\,+}\right),D_{-\frac{5}{6}}^++D_\frac{1}{6}^{d\,-},i\left(D_{-\frac{5}{6}}^+-D_\frac{1}{6}^{d\,-}\right),\sqrt{2}S_{-\frac{1}{3}}\right)^T\,,
\eeq
where the $\pm$ superscripts denote the $T_L^3=\pm 1/2$ components of the corresponding SU(2)$_L$ doublet. 
$\lag_{\rm elem}$ is invariant under a global $[$SU(2)$_L\times$U(1)$_\mathcal{Y}]^{\rm el}$ under which $q_L$ and $u_R$ $(d_R)$ transform as a doublet and a singlet with U(1) charge $1/6$ and $2/3$ $(-1/3)$, respectively.
$\lag_{\rm mix}$  explicitly breaks  the global symmetries of the elementary and strong sectors. Yet, the mixings in $\lag_{\rm mix}$ are the most general terms which are linear in the fermion fields and respect the diagonal subgroup $[$SU(2)$_L\times$U(1)$_\mathcal{Y}]^{\rm SM}$, which is gauged and identified with the SM EW gauge group. $D_\mu$ denote the SM covariant derivative. 
 
Despite being independent parameters, the composite resonance masses do not generically display hierarchies in models where the strong sector is characterized by a single scale and thus $M_u\sim M_d$ is expected. For simplicity we will assume in the following degenerate up and down masses, $M_u=M_d\equiv M$, in order to simplify the algebra and obtain closed expressions, but the physical results will be independent of this assumption.
Given the Lagrangian in Eq.~\eqref{LMCHM}, the spectrum of the heavy resonances goes as follows. Prior to EWSB and in the absence of elementary/composite mixing, the {\bf 4} and the {\bf 1} components of $\Psi^i$, $i=u,d$, are eigenstates of mass $M$ and $M+Y_if$, respectively. The singlet thus receives an SO(5) breaking contribution from the strong sector Yukawa and can be heavier or lighter than the fourplet depending on the sign of $Y_i$. Switching on the elementary/composite mixings, the singlet eigenmasses become $M_S^i= ((M+Y_i f)^2+\lambda_i^2)^{1/2}$, and one linear combination of the doublets mixing with $q_L$ receives an extra contribution from mixing and its mass becomes $M_D=(M^2+\lambda_{q^u}^2+\lambda_{q^d}^2)^{1/2}$, while masses of the unmixed doublets remain unchanged (we have assumed here and for the rest of the paper that the mixing masses $\lambda_i$ are real). EWSB will further mix states of same electric charge, thus yielding $\mathcal{O}(v/f)$ splitting in the above spectrum.
More precisely, the mass matrices of the $Q=2/3$ states reads
\beq
-\lag_{\rm mass}^{Q=2/3}= \bar\psi^u_{L}\mathcal{M}_{u}\psi^u_{R}+\hc\,,\nonumber
\eeq
\beq
\label{eq:massmatrix5}
 \mathcal{M}_{u}=\left(\begin{array}{ccccc}
0& \lambda_{q^u} & 0 & 0 & \lambda_{q^d} \\
0 & M +\frac{Y_u f}{2} \sin^2\frac{h}{f}& \frac{Y_u f}{2} \sin^2\frac{h}{f} & \frac{Y_u}{2\sqrt{2}} f \sin\frac{2h}{f} & 0\\
0 & \frac{Y_u f}{2} \sin^2\frac{h}{f} & M+\frac{Y_u f}{2} \sin^2\frac{h}{f} & \frac{Y_u}{2\sqrt{2}} f \sin\frac{2h}{f} & 0\\
\lambda_u & \frac{Y_u}{2\sqrt{2}} f \sin\frac{2h}{f} & \frac{Y_u}{2\sqrt{2}} f \sin\frac{2h}{f} & M+Y_u f\cos^2\frac{h}{f} & 0 \\
0 & 0 & 0 & 0 & M
\end{array}\right)
\eeq
in a basis where $\psi^u_{L}=(q^+_L,D^{u+}_{\frac{1}{6}L},D^{u-}_{\frac{7}{6}L},S_{\frac{2}{3}L},D^{d+}_{\frac{1}{6}L})^T$ and $\psi^u_{R}=(u_R,D^{u+}_{\frac{1}{6}R},D^{u-}_{\frac{7}{6}R},S_{\frac{2}{3}R},D^{d+}_{\frac{1}{6}R})^T$.

Finally, in unitary gauge, {\it i.e.} removing the EW Goldstone bosons, the $\Sigma$ field takes the form (see Appendix~\ref{Sigma})
\beq
\Sigma=\left(0,0,\sin \frac{h}{f},0,\cos \frac{h}{f}\right)\,,
\eeq
where $h$ is the physical Higgs component, with $\langle h \rangle = v\neq 0$. Although  generically $v\sim f$, the Higgs VEV is to be mildly tuned in order to agree with various EW precision measurements from LEP~\cite{AgasheContino,SILH} (generating a 125~GeV mass for the Higgs boson also requires some mild tuning~\cite{Panico:2012uw,14in5D}). Thus $v/f\lesssim 0.5$ is expected and Higgs non-linearity effects are well enough captured at leading order by the dimension six operators of Eq.~\eqref{LNP}.\\

We match now $\lag_{\rm 2site}$ to $\lag_{\rm eff}$, beginning with
pure Higgs operators. Expanding the two-derivative Lagrangian of $\Sigma$ 
and matching the Higgs kinetic term and the $W$ mass to the EFT yields\footnote{In the SILH basis~\cite{SILH} where $c_r=0$, one finds  $c_H^{\rm SILH}=1/f^2$ and $c_y^{\Sigma\,{\rm SILH}}=1/f^2$, see footnote\footnoterecall{cr}.}
\beq\label{crH}
c_H = -\frac{1}{2}c_r = \frac{1}{3f^2}\,.
\eeq

Then, integrating out the heavy resonances at tree-level one finds (neglecting flavor violation) 
\beq
y_u= Y_u\sin\theta_{q}\cos\phi\sin\theta_u\,,\quad y_d=Y_d \sin\theta_q\sin\phi\sin\theta_d\,,
\eeq
where 
\beq
\sin\theta_{q}\equiv \frac{\lambda_{q}}{\sqrt{\lambda_{q}^2+M^2}}\,,\quad
\sin\theta_{i}\equiv \frac{\lambda_{i}}{\sqrt{\lambda_i^2+(M+Y_if)^2}}\,,
\eeq
are the sines of the LH and RH mixing angles, respectively, $\tan\phi \equiv \lambda_{q^d}/\lambda_{q^u}$, $\lambda_q=\sqrt{\lambda_{q^u}^2+\lambda_{q^d}^2}$, and 
\beq
c_{y_i}\equiv c_y^{\Sigma}+c_{y_i}^{\Psi}\quad {\rm with }\quad c_y^{\Sigma}=\frac{4}{3f^2}\,,\quad
 c_{y_i}^{\Psi}=\sin^2\theta_i\frac{Y_i(2M+Y_if)}{fM^2}+\mathcal{O}(\sin\theta_{q}^2)
\,,\label{cyMCHM5}
\eeq
for $i=u,d$. It can be checked that the light quark mass given by~Eq.~\eqref{masses} is indeed an eigenvalue of the mass matrix~\eqref{eq:massmatrix5} at the order $\mathcal{O}(v^4)$. Again, we do not consider composite LH quarks, since it is strongly disfavored by LEP, hence we assumed $\sin\theta_{q}\ll 1$ and neglected $\mathcal{O}(\sin\theta_{q}^2)$ effects. (We provide  nonetheless the complete expressions of $c_y^\Psi$ in Appendix~\ref{1014matching}.)
$c_y^{\Sigma}$ is the contribution from pure Higgs non-linearities, while $c_y^\Psi$, which decouples with $M\to \infty$, arises from the presence of heavy fermionic resonances.\footnote{Notice that although $c_y^{\Sigma}$ is independent on the mass and mixings of the associated resonances, it does depend on their quantum numbers under the strong sector global symmetries, see Appendix~\ref{Irreps} for some other explicit exemples.} 
Notice that $c_{y_i}^\Psi$ vanish in the limit of zero mixing $\sin\theta_{q}=\sin\theta_i=0$, as expected from the exact Goldstone symmetry of the strong dynamics.\footnote{This result becomes explicitly clear upon going to the unitary gauge, by mean of an appropriate SO(5) transformation, in which  the NGBs are completely removed from the Yukawa operator in $\mathcal{L}_{\rm strong}$~\cite{AzatovGal}.} 

One is then left with the determination of $c_{g,\gamma}$ through one-loop matching of the $gg\to h$ and $h\to \gamma\gamma$ amplitudes. In order to do so, we take the formal limit where the SM-like quarks are heavier than the Higgs boson ($m_{u,d}/m_h\to \infty$) and rely on LEHT to match for $c_{g,\gamma}$. One could also explicitly evaluate the one-loop diagrams relevant for this amplitude matching. Yet, since $c_{g,\gamma}$  are controlled by the NP scale, their matching values do not depend on the SM masses and the use of the LEHT, which is legitimate in the heavy mass limit, is much more practical. For instance the $gg\to h$ EFT amplitude Eq.~\eqref{Mhgg} becomes in the limit $\tau_{u,d}\to 0$
\beq\label{MhggEFT}
\mathcal{M}_{gg\to h}&\propto& c_{g}v^2 +\sum_{i=u,d} \left[1- v^2\left({\rm Re}[c_{y_i}]+\frac{c_H}{2}\right)\right]\,,
\eeq
while in the SO(5)$/$SO(4) model the LEHT yields 
\beq\label{MhggPGB}
\mathcal{M}_{gg\to h}^{{\rm MCHM}}\propto \partial_{\,\log v}\log\det\mathcal{M}(v)=2-3\xi+\mathcal{O}(\xi^2)\,,
\eeq
where $\xi\equiv v^2/f^2$, $\mathcal{M}(v)$ is the Higgs background dependent mass matrix of the $Q=2/3$ and $Q=-1/3$ fermions, whose determinant factorizes (since $Q$ is conserved) as $\det\mathcal{M}=\det\mathcal{M}_u\times \det\mathcal{M}_d$, with $\det\mathcal{M}_i (v) =Y_ifM\lambda_i\lambda_{q^i}\sin(v/f)\cos(v/f)/\sqrt{2}$ and $v= f\arcsin(f/v_{\rm SM})$.  Again, since the resonance effects cancel out in the heavy mass limit, Eq.~\eqref{MhggPGB} is only driven by Higgs non-linearities. Finally, matching the amplitudes Eq.~\eqref{MhggEFT} and Eq.~\eqref{MhggPGB}, together with using the tree-level results Eqs.~\eqref{crH} and \eqref{cyMCHM5}, yields
\beq
c_g = \sum_{i=u,d} {\rm Re}[c_{y_i}^\Psi]= \sum_{i=u,d} \sin^2\theta_i\frac{Y_i(2M+Y_if)}{fM^2}+\mathcal{O}(\sin^2\theta_{q})\,.\label{cgMCHM5}
\eeq
A similar derivation for $h\to \gamma\gamma$ gives 
\beq\label{cgammaMCHM5}
c_\gamma = \sum_{i=u,d} Q_i^2 {\rm Re}[c_{y_i}^\Psi]\,.
\eeq
Hence, for an heavy quark like the top, the effects of the strong dynamics on radiative Higgs couplings is driven by Higgs non-linearities and {\it e.g.}
\beq
\mathcal{M}_{gg\to h}^{m_u\gg m_h}\propto 1- \left(c_y^\Sigma+\frac{c_H}{2}\right)v^2=1-\frac{3}{2}\xi\,,
\eeq
while for a light flavor, Higgs couplings are only shifted by $c_{g,\gamma}$, {\it e.g.}
\beq
\mathcal{M}_{gg\to h}^{m_{u,d}\ll m_h}\propto c_{g}v^2\,,
\eeq
which is negligible unless the RH chirality is relatively composite. 

We study  in Section~\ref{sec:pheno} the impact of the above effects on Higgs physics at hadron colliders.

\section{Composite Flavor Physics}\label{sec:flavor}
The two-site description of composite Higgs models is somewhat limited when one considers flavor physics since the generic new physics scale probed by flavor precision observables is as high as  few thousands of TeV~\cite{INP}, which is well above the effective cutoff of the composite sector ($\sim\,$TeV).
However, one can view the two-site picture as an effective description of a more complete theory of flavor inspired by holography~\cite{holog} in which order one anomalous dimensions for chiral operators would induce the large SM flavor hierarchies~\cite{NeuGross,Huber,GhergPom}. 

We first briefly recall the benefits of such a theory with regard to flavor physics and contrast it with Froggatt--Nielsen (FN) type of theories~\cite{FN} in which the SM flavor hierarchies arise from $\mathcal{O}(1)$ different charges of the different generations under an additional global U(1) horizontal symmetry. Then, we review the essence of various flavor constructions in the complete microscopic theory and describe the resulting structures for two-site model flavor parameters.

\subsection{Strong dynamics vs. abelian flavor symmetries}

As far as only the structure of the SM Yukawas is concerned, the flavor structure of the microscopic composite Higgs theory looks very similar to  those obtained from FN constructions, like in split fermion models within flat extra dimension~\cite{Arkani-Hamed-Schmaltz,Grossman:2004rm}. However, one major difference lies in the way SM fermions couple to new physics fields, like scalar quarks (squarks) in supersymmetric models or gauge Kaluza--Klein (KK) excitations in extra-dimensional models. 
The reason is fairly simple. In FN models diagonal entries of NP flavor parameters, like the squark mass squared matrix in SUSY or the KK-gluon to SM fermions couplings in extra-dimensions, are invariant under the horizontal symmetries. Thus they can all be of the same order, which generically yields overly large flavor violation effects for the first two generations. 
Conversely, in models where the SM flavor hierarchies are obtained from set of sizable (random) anomalous dimensions, the contribution to the diagonal entries of the NP flavor parameters are hierarchical and exponentially suppressed for the first two generations. This is the reason why models of abelian flavor symmetries are subject to stricter constraints from flavor observables related to the first two generations than strong dynamics models based on large anomalous dimensions.
Although this mechanism is inherent to models of strong dynamics (or warped extra-dimensions) with partial compositeness, an implementation in SUSY is possible as in Nelson--Strassler models~\cite{NelsonStrassler}.

\subsection{Composite flavor structures}
In holographic dual descriptions of models of strong dynamics, the microscopic (fundamental) flavor parameters are the five dimension (5D) fermion masses and the 5D Yukawa couplings, which are respectively dual to the large anomalous dimensions and the inter-composite Yukawas in 4D strongly coupled theories.
All existing studies on the flavor structure of such models fall into three broad classes. We describe below how their respective assumption on the microscopic flavor parameters differ, as well as the flavor structures they match onto in the two-site effective description used in the paper.

\begin{itemize}
\item Class (I) {\it Anarchy}: All fundamental flavor parameters are structureless, {\it i.e.}  anarchic. 
This is the most explored case so far. It consists of an appealing integral mechanism to generate SM flavor hierarchies~\cite{NeuGross,Huber,GhergPom}, where SM mass hierarchies are dictated by the relative degree of compositeness of SM fermions. In the two-site picture, heavy SM fermions like the top quark are thus interpreted as mostly composite objects ($\epsilon\sim\mathcal{O}(1)$), while lighter SM fermions are mostly elementary fields ($\epsilon\ll 1$). Most importantly the same integral mechanism also protects the model against large contributions to flavor-changing neutral current (FCNC) processes through a GIM-like mechanism~\cite{HuberFV,APS}. However, this so-called RS--GIM mechanism is not perfect and overly large (CP violating) contributions to FCNCs in the down sector as well as to electric dipole moments are generically induced~\cite{APS,CsakiFalkowskiWeiler,KaonCP}.
It is worth recalling though that SM flavor hierarchies together with a similar suppression for flavor violating processes can be obtained in anarchic models where the hierarchy problem is only solved up to scale much lower than the Planck scale~\cite{LRS}. 
The anarchic extra dimension model matches onto a two-site model where the composite Yukawas are anarchical but the elementary/composite mixings are hierarchical and quasi-aligned with the SM Yukawa matrices~\cite{APS,NMFV}.

\item Class (II) {\it Minimal flavor violation} (MFV): The microscopic flavor parameters are hierarchical and realize the 4D MFV selection rules~\cite{MFV,GMFV}. The SM flavor puzzle remains unsolved but the theory entertains a strong mechanism to suppress new sources of flavor breaking~\cite{RZ,5DMFV,CsakiWeiler,FTRS,RediWeiler}.
The literature on this class of models can be divided into two subclasses:
(IIa) Flavor triviality~\cite{FTRS, FTRS2}: the anomalous dimensions, as well as the Yukawas of the microscopic theory are proportional to SM Yukawas. As a consequence the composite site Yukawas are also proportional to the SM Yukawas while the mass terms mixing the two sites are degenerate for the first two generations, but generically split from the third one~\cite{GMFV}.
(IIb) Composite universality~\cite{RediWeiler,Redi}: The  microscopic theory is invariant under one or several U(3) vectorial flavor symmetries. Hence, this results in two-site composite Yukawas along with some of the elementary/composite mixings which are proportional to the identity matrix, while the remaining mixings are proportional to the SM Yukawas. 

\item Class (III) {\it Exhilaration}: The anomalous dimensions are anarchic, yet it is  possible for the first two generation quarks to be composite. The microscopic Yukawas may result being partially hierarchical~\cite{Exhil}. This case is subject to severe flavor violation constraints, so some additional mechanism of alignment, through {\it e.g.} using horizontal symmetries, has to be implemented. The corresponding two-site model flavor parameters consists of composite Yukawas and elementary/composite mixings which are also partially hierarchical.
\end{itemize}

It will be useful in the remainder of the paper to treat separately the ``top sector'', consisting of the LH and RH top and the LH bottom quarks, which is expected to be composite in order to accomodate the large top mass, from the remaining ``light quark sector'', whose level of compositeness is model dependent. 
As we argued above, one does expect  mostly elementary light quarks in class (I), whereas in classes (II) and (III) some of the light quarks could be composite without conflicting with precision flavor observables.
Moreover, for class (II) models, one expects either the first two (case IIa) or all three (case IIb) generations to have degenerate flavor parameters as a result of the corresponding U(2) or U(3) flavor symmetries.

\section{Phenomenological Implications} \label{sec:pheno}

We study now in greater details the implications of composite light quarks on Higgs rates at the LHC. For definiteness we focus on MCHM based on the SO(5)$/$SO(4) coset and where composite fermions are embedded into fundamentals of SO(5), 
but our results can be straightforwardly extended to less minimal fermionic sector. In order to remain consistent with EW precision measurements we assume that only RH quarks can be sizably composite~\cite{FTRS,PontonSerone,RediWeiler}. 
The net effect of strong sector resonances on Higgs couplings depends on the number of composite flavors and their respective degree of compositeness. We do not commit to any specific flavor setup but simply assume below that $N_u$ ($N_d$) RH light up (down) flavors can be significantly composite. We will always assume RH bottoms to be mostly elementary to keep emphasis on first two generation effects (see Refs.~\cite{AzatovGal,AzatovGalreview} for a discussion of composite RH bottom). Thus, we have $N_{u,d}\leq 2$. We will further assume degenerate flavor parameters whenever more than one generation is taken significantly composite, which is a natural prediction of class (II) models realizing the MFV ansatz. Predictions from other (non-degenerate) scenarios can easily be derived as well.

Fermionic resonances associated with composite light generation quarks impact  Higgs physics dominantly through couplings of the Higgs to gluons and photons. Therefore we focus on the Higgs signal strengths where the above effects are more pronounced, that is in the $\gamma\gamma$ channel, and in the $ZZ^*$ and $WW^*$ channels since most of these events are produced from gluon-fusion. We do not consider $b\bar b$ final states since those are only observable at the LHC through $W/Z$ associated production.  

 Higgs signal strengths $\mu_i$ are defined as the product of the production cross-section times the branching ratio into final states $i=\gamma\gamma$, $ZZ^*$ and $WW^*$ relative to the SM ones, {\it i.e.}
\beq
\mu_{i}=\frac{\sum_j\sigma_{j\to h}\times {\rm Br}_{h\to i}}{\sum_j\sigma_{j\to h}^{\rm SM}\times {\rm Br}_{h\to i}^{\rm SM}}\,,
\eeq
where $j$ runs over all Higgs production modes, by far the dominant one being gluon fusion. The vector boson fusion (VBF) production cross-section is modified at tree-level due to the non-linear nature of the Higgs and also potentially by the presence of light spin one resonances. Given the present $\mathcal{O}(1)$ uncertainty in VBF tagged diphoton rate and the smallness of the later relative to the untagged rate, we will only consider corrections to the gluon fusion cross-section. Assuming gluon fusion dominance, signal strengths factorize as 
\beq
\mu_i\simeq X_{gg}\times R_i\,,
\eeq
where we defined $X_{gg}\equiv \sigma_{gg\to h}/\sigma_{gg\to h}^{\rm SM}$ as the gluon fusion production cross-section ratio and $R_i\equiv {\rm Br}_{h\to i}/{\rm Br}_{h\to i}^{\rm SM}$ as the branching ratio into the final states $i$ ratios.

\subsection{Higgs Production}

\begin{figure}[t]
\centering
\begin{tabular}{cc}
\includegraphics[scale=0.55]{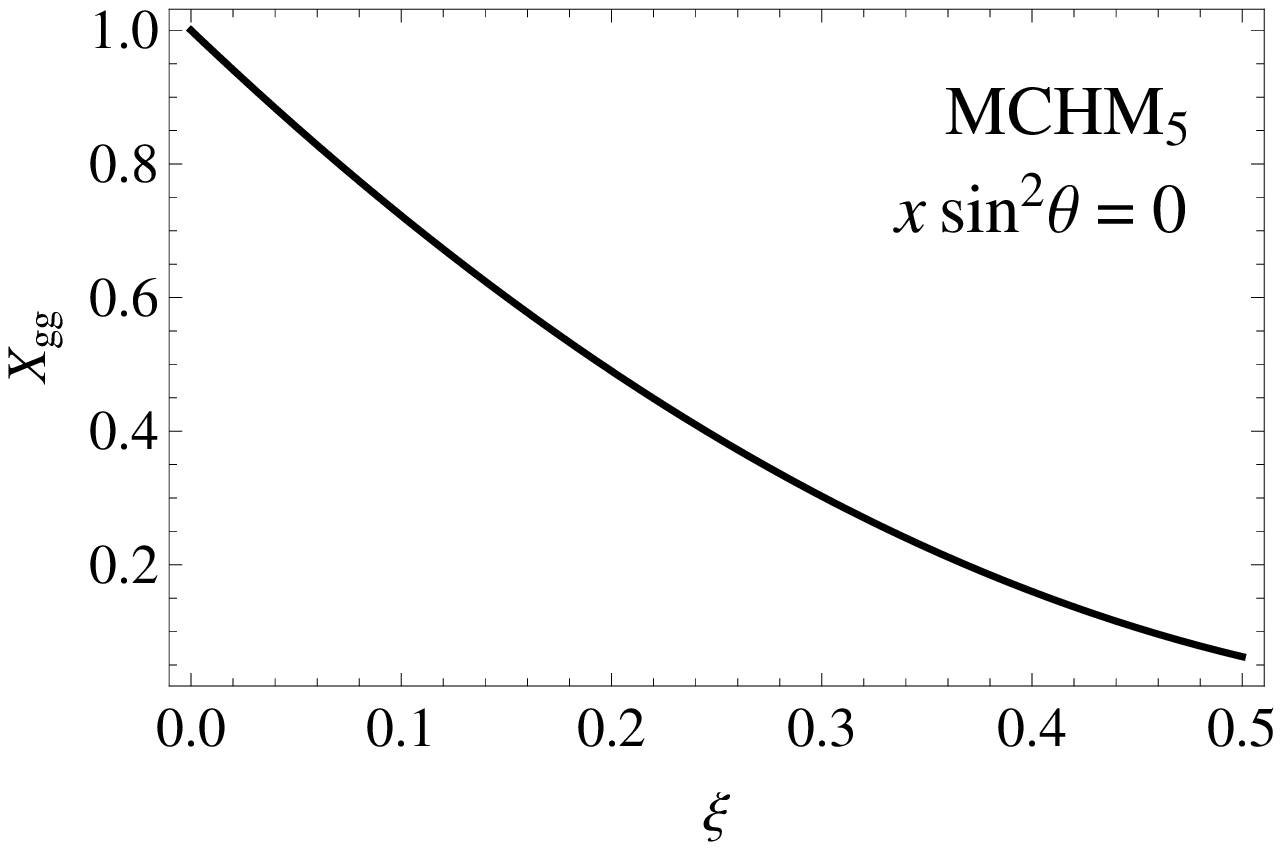}&\includegraphics[scale=0.55]{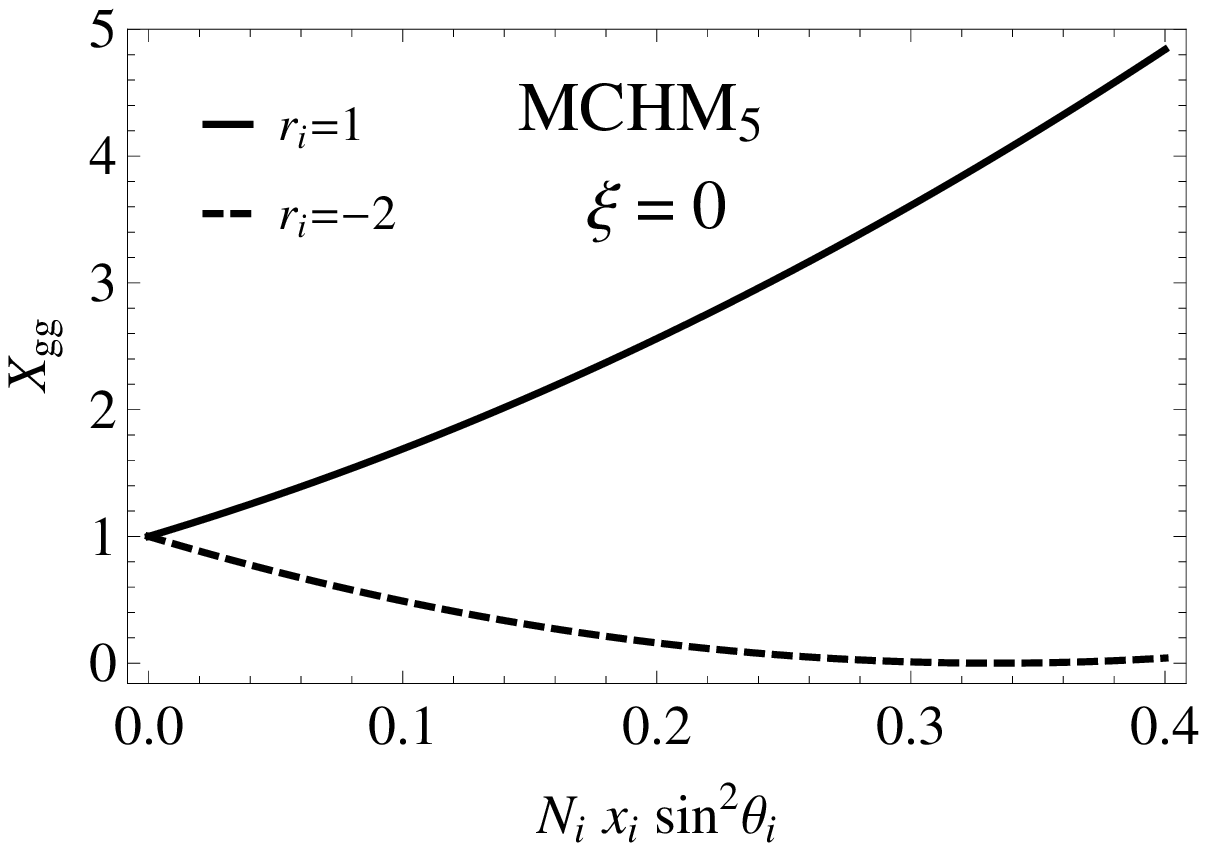}
\end{tabular}
\caption{$X_{gg}$ ratio of gluon fusion Higgs production cross-section in MCHM relative to SM as a function of $\xi$ (setting $N_ix_i \sin^2\theta_i=0$) [left] and $N_ix_i \sin^2\theta_i$ (setting $\xi=0$) [right], for $i=u$ or $d$. We defined $\xi=v^2/f^2$, $x_i=(Y_iv/M_i)^2$ and $r_i=M/(Y_if)$.}
\label{fig:Rggindiv}
\end{figure}

\begin{figure}[t]
\centering
\begin{tabular}{cc}
\includegraphics[scale=0.5]{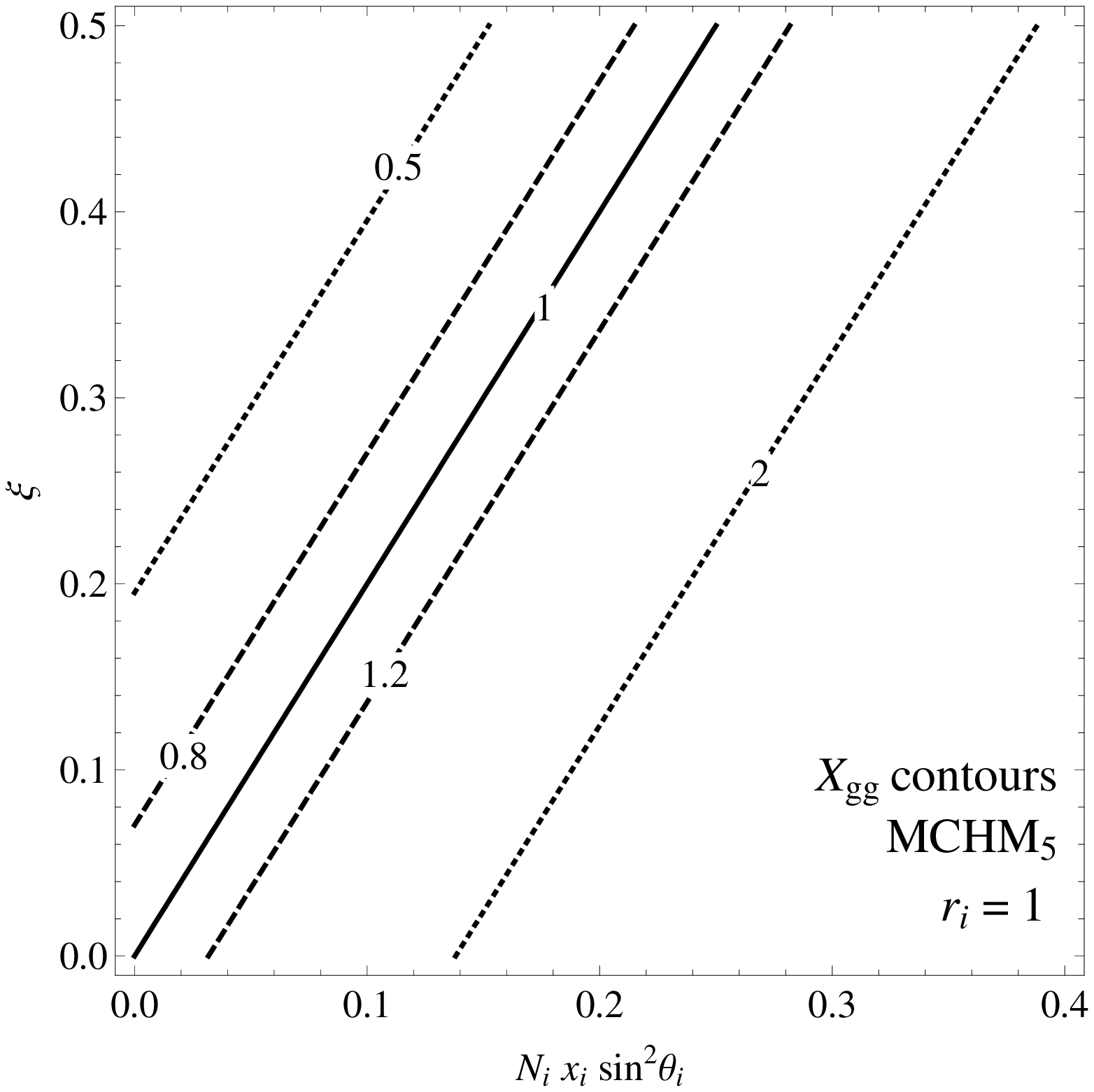}&\includegraphics[scale=0.5]{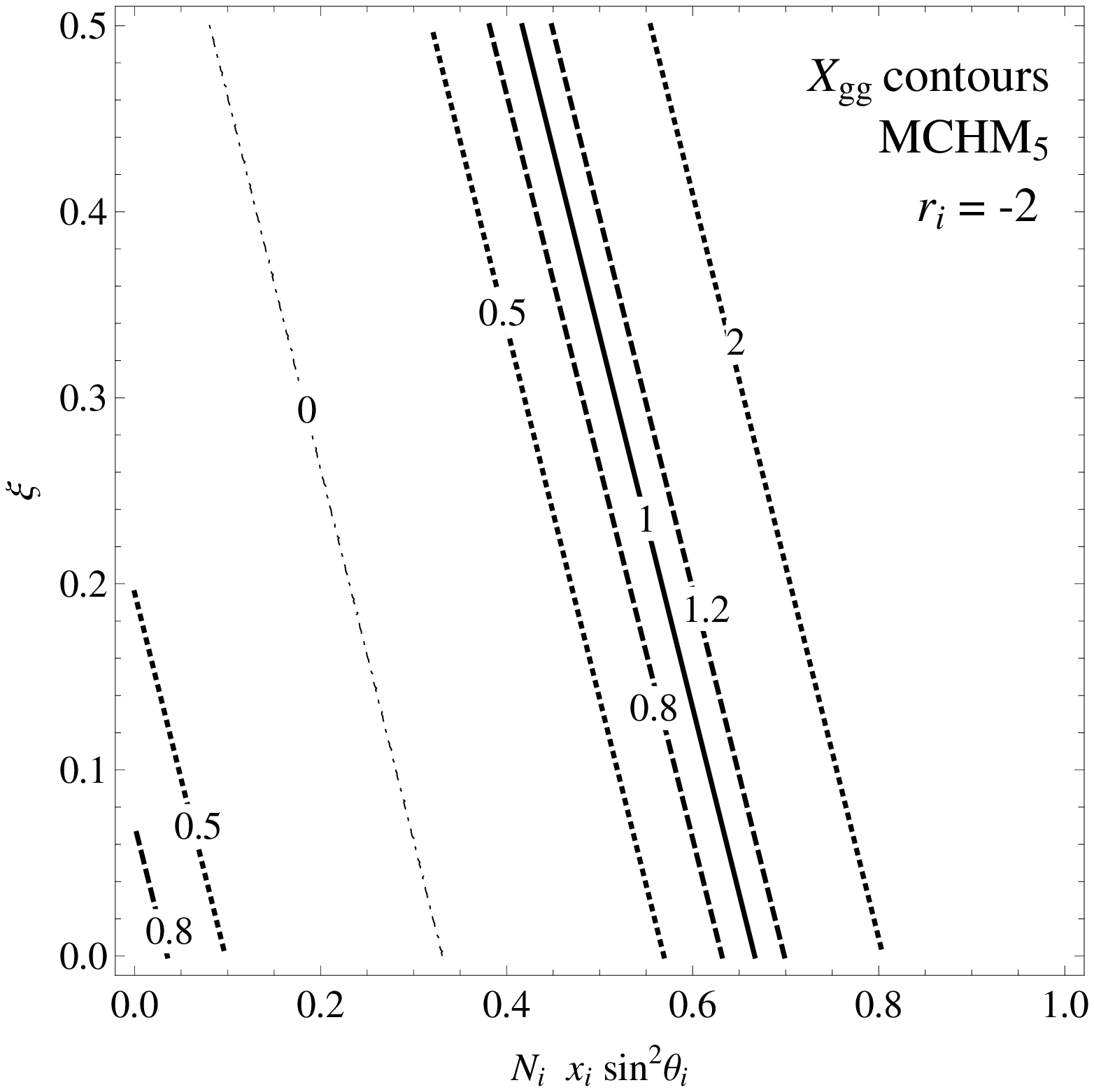}
\end{tabular}
\caption{$X_{gg}$ ratio of gluon fusion Higgs production cross-section in MCHM for $N_i$ ($i=u$ or $d$) RH composite flavors as a function of $\xi= v^2/f^2$, the RH elementary/composite mixing $\sin\theta_i$ and $x_i=(Y_iv/M)^2$. Red (black) contours correspond to enhancement (suppression) relative to the SM cross-section.}
\label{fig:Rgg5}
\end{figure}

From Eqs.~\eqref{Mhgg},~\eqref{cyMCHM5} and \eqref{cgMCHM5}, we find,  to leading order in $\tau_t=m_h^2/(4m_t^2)$,  the following contributions to $X_{gg}$ in MCHM
\beq\label{Xgg5}
X_{gg}^{{\rm MCHM}} \simeq 1-3\xi +2\sum_{i=u,d} N_i x_i\sin^2\theta_i \left(1+2r_i\right)+\dots\,,
\eeq
where we introduced the dimensionless parameters
\begin{equation} 
x_i\equiv (Y_iv/M)^2 \ \ \ \textrm{and} \ \ \ r_i\equiv g_\Psi/Y_i,
\end{equation}
with $g_\Psi \equiv M/f\lesssim 4\pi$ a fermionic strong coupling constant, and $\dots$ denotes higher orders in $\xi$ and $x_i$. If all fermion couplings are of comparable size we expect $r\sim \mathcal{O}(1)$ and $x=(v/f)^2(Y/g_\Psi)^2\sim \mathcal{O}(\xi)$. Note that the sign of $r$ is not fixed. 
The first new physics term in Eq.~\eqref{Xgg5} is the effect of the top sector. It is only controlled by Higgs non-linearities, due to the aforementioned cancelation, and lead to a suppressed production cross-section through gluon fusion~\cite{AdamHgg,LowVichi}. Note that there is no contribution from the composite LH bottom when it mixes with a ${\bf 5}$ representation~\cite{AzatovGal}. Although the top sector contribution is insensitive to the top partners spectrum (and to the top compositeness), it does depend on their representation under the strong sector symmetries. Nonetheless, as we show in Appendix~\ref{1014matching} the $\mathcal{O}(\xi)$ contribution to $X_{gg}$ also leads to a suppressed Higgs production cross-section for top partners in the ${\bf 10}$ or the ${\bf 14}$ representation of SO(5).\footnote{This is in fact generically expected in NP models where the Higgs boson is naturally light~\cite{LowRattazzi}.} The last term in Eq.~\eqref{Xgg5} is the contribution from strong sector partners of the RH light quarks, which can either enhance or further suppress the gluon fusion cross-section, depending on the sign of $1+2r$. We show the impact on $X_{gg}$ of each term separately in Fig.~\ref{fig:Rggindiv}.
When both effects are present there is a region of parameters where  they balance each other and where, as shown in  Fig.~\ref{fig:Rgg5}, $X_{gg}\simeq 1$ is achieved without decoupling the scale of the strong dynamics (as would be required for $x\sin^2\theta=0$), even for a single composite RH quark. For elementary RH light quarks, Higgs non-linearities yield a large suppression of the gluon fusion cross-section of ${\it e.g}$ $\sim 50\%$ for a moderately small $\xi\simeq 0.2$ ($f\simeq 550\,$GeV). On the other hand, if one or several RH light quarks are relatively composite objects, large enhancements are expected up to a factor of a few. Note that 
when $r<-1/2$ the resonance contribution interferes destructively with the SM one, which thus leads, as shown on the right panel of Fig.~\ref{fig:Rgg5}, to either a completely suppressed or largely enhanced gluon fusion cross-section, depending on the value of $x \sin^2\theta$.

\subsection{Higgs decay widths}

We move now to consider Higgs decays into gauge bosons. From Eqs.~\eqref{Mhyy}, ~\eqref{Mhothers},~\eqref{cyMCHM5} and ~\eqref{cgammaMCHM5} we find the following correction in MCHM to the $h\to \gamma\gamma$  branching ratio
\beq
R_{\gamma\gamma}^{{\rm MCHM}}\simeq\frac{1}{1-\delta}\left[1+\mathcal{A}_{\rm SM}^{-1}\left[\left(\frac{7}{4}A_1(\tau_W)-3Q_u^2\right)\xi
+2\sum_{i=u,d}N_i Q_i^2x_i\sin^2\theta_i (1+2r_i)\right]+\dots\right]\,,\label{Ryy5}
\eeq
where $\mathcal{A}_{\rm SM}\equiv Q_u^2-\frac{7}{4}A_1(\tau_W)\simeq -1.6$, and to the $h\to WW^*,ZZ^*$ branching ratio
\beq
R_{WW^*}^{{\rm MCHM}}&=&R_{ZZ^*}^{{\rm MCHM}}= \frac{1}{1-\delta}\left[1-\xi+\mathcal{O}(\xi^2)\right]\,,\label{Rzz5}
\eeq 
while
\beq
R_{gg}^{{\rm MCHM}}= \frac{X_{gg}^{\rm MCHM}}{1-\delta}\, , \ \ \
R_{bb}^{\rm MCHM} = \frac{1}{1-\delta} \left[ 1-3\, \xi + \mathcal{O}(\xi^2)\right] \,.\label{Rbb5}
\eeq 
Note that the $WW^*$ and $ZZ^*$ branching ratios receive the same correction thanks to custodial symmetry.

$\delta\equiv 1-\Gamma_h^{\rm MCHM}/\Gamma_h^{\rm SM}$ captures the correction to the branching ratios due to a change in the total Higgs width $\Gamma_h$, relative to the SM.
It is convenient to write it as 
\beq
\delta = \sum_i {\rm Br}_{h\to i}^{\rm SM} \times \left(1 -\gamma_i\right)\,, \quad \gamma_i \equiv \frac{\Gamma_{h\to i}^{\rm MCHM}}{\Gamma_{h\to i}^{\rm SM}}\, ,
\eeq
where $\Gamma_{h\to i}^{\rm SM}$ and $\Gamma_{h\to i}^{\rm MCHM}$ are the partial decay widths for the channel $i$ in the SM and MCHM, respectively. We only take into account the decay channels $i=b\bar b$, $WW^*$, $gg$ and $ZZ^*$ which dominate the total width for a 126$\,$GeV Higgs boson and for which the corresponding SM branching ratios are ${\rm Br}_{h\to b\bar b}\simeq 60\%$, ${\rm Br}_{h\to WW^*}\simeq 20\%$, ${\rm Br}_{h\to gg}\simeq 10\%$ and ${\rm Br}_{h\to ZZ^*}\simeq 3\%$~\cite{HXSWG}. 
From Eq.~\eqref{Mhothers} we find 
\beq
\gamma_{b\bar b} = 1-3\xi+\mathcal{O}(\xi^2)\,,\quad
\gamma_{WW^*}= \gamma_{ZZ^*} = 1-\xi+\mathcal{O}(\xi^2)\,, 
\eeq
while $\gamma_{gg}= X_{gg}^{{\rm MCHM}}$. Note again that when the LH bottom mixes with a ${\bf 5}$ representation of the strong sector, $\gamma_{b\bar b}$ is insensitive to the LH bottom compositeness~\cite{AzatovGal}. Thus, under the assumption of an elementary RH bottom quark the $h\to b\bar b$ coupling is only modified by Higgs non-linearities through a flavor universal $c_y^\Sigma$ contribution.
Plugging back the above expressions for $\gamma_i$ into Eqs.~\eqref{Ryy5},~\eqref{Rzz5} and~\eqref{Rbb5} yields
\beq
R_{\gamma\gamma}^{\rm MCHM} &\simeq & 1+1.9\, \xi -0.2\sum_{i=u,d} N_ix_i\sin^2\theta_i(1+2r_i)(1+6.1 Q_i^2)+\dots\,\\
R_{WW^*}^{\rm MCHM}=R_{ZZ^*}^{\rm MCHM}&\simeq & 1+1.3\, \xi -0.2\sum_{i=u,d} N_ix_i\sin^2\theta_i(1+2r_i)+\dots\,\\
R_{gg}^{\rm MCHM} & \simeq &  1 - 0.7 \, \xi +1. 8 \sum_{i=u,d} N_ix_i\sin^2\theta_i(1+2r_i)+\dots\, \\
R_{bb}^{\rm MCHM} & \simeq &  1 - 0.7 \, \xi +0. 2 \sum_{i=u,d} N_ix_i\sin^2\theta_i(1+2r_i)+\dots\, ,
\eeq 
where the $\dots$ denote higher orders in $\xi$ and $x$. Therefore, pure Higgs non-linearities lead to an enhancement in the branching ratios in diphotons and weak bosons, which is incidentally of comparable size. On the other hand, light RH quark compositeness tends to suppress (enhance) the latter for $r>-1/2$ ($r<-1/2$).

\subsection{Signal strength into photons and weak bosons}

\begin{figure}[t]
\centering
\begin{tabular}{cc}
\includegraphics[scale=0.55]{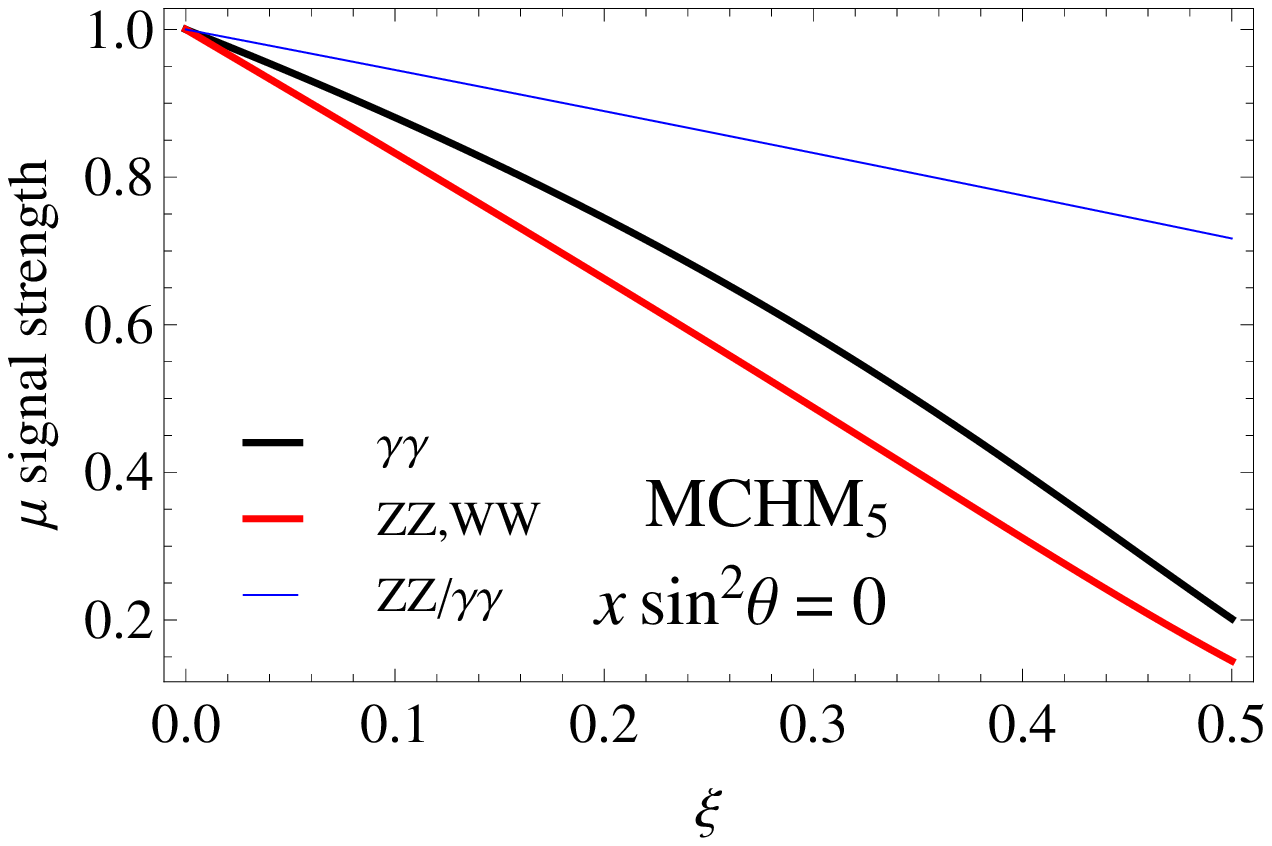}&\includegraphics[scale=0.55]{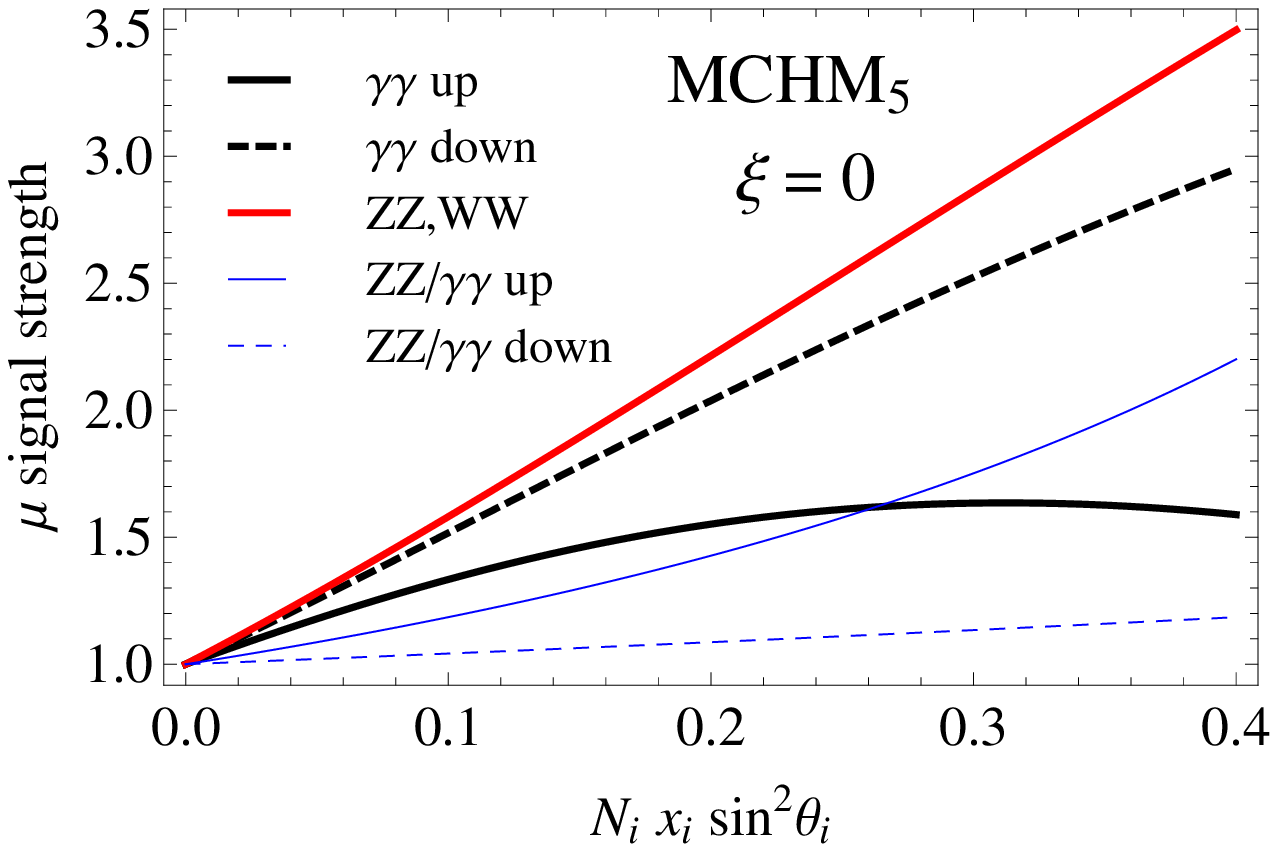}
\end{tabular}
\caption{Higgs signal strengths into EW gauge bosons as a function of $\xi$ (setting $N_i x_i \sin^2\theta_i=0$) [left] and $N_i x_i \sin^2\theta_i$ (setting $\xi=0$) [right] in MCHM; $i=u$ or $d$. $s_i$ is the RH elementary/composite mixing and we defined  $\xi = v^2/f^2$, $x_i=(Y_iv/M)^2$ and $r_i=M/(Y_if)=1$. For the diphoton signal strength, we considered two cases where either $N_u$ RH up-type or $N_d$ down-type quark flavors are composite. Blue lines show the $\mu_{ZZ,WW}/\mu_{\gamma\gamma}$ signal strength ratio.}\label{mu5split}
\end{figure}

\begin{figure}[t]
\centering
\begin{tabular}{cc}
\includegraphics[scale=0.5]{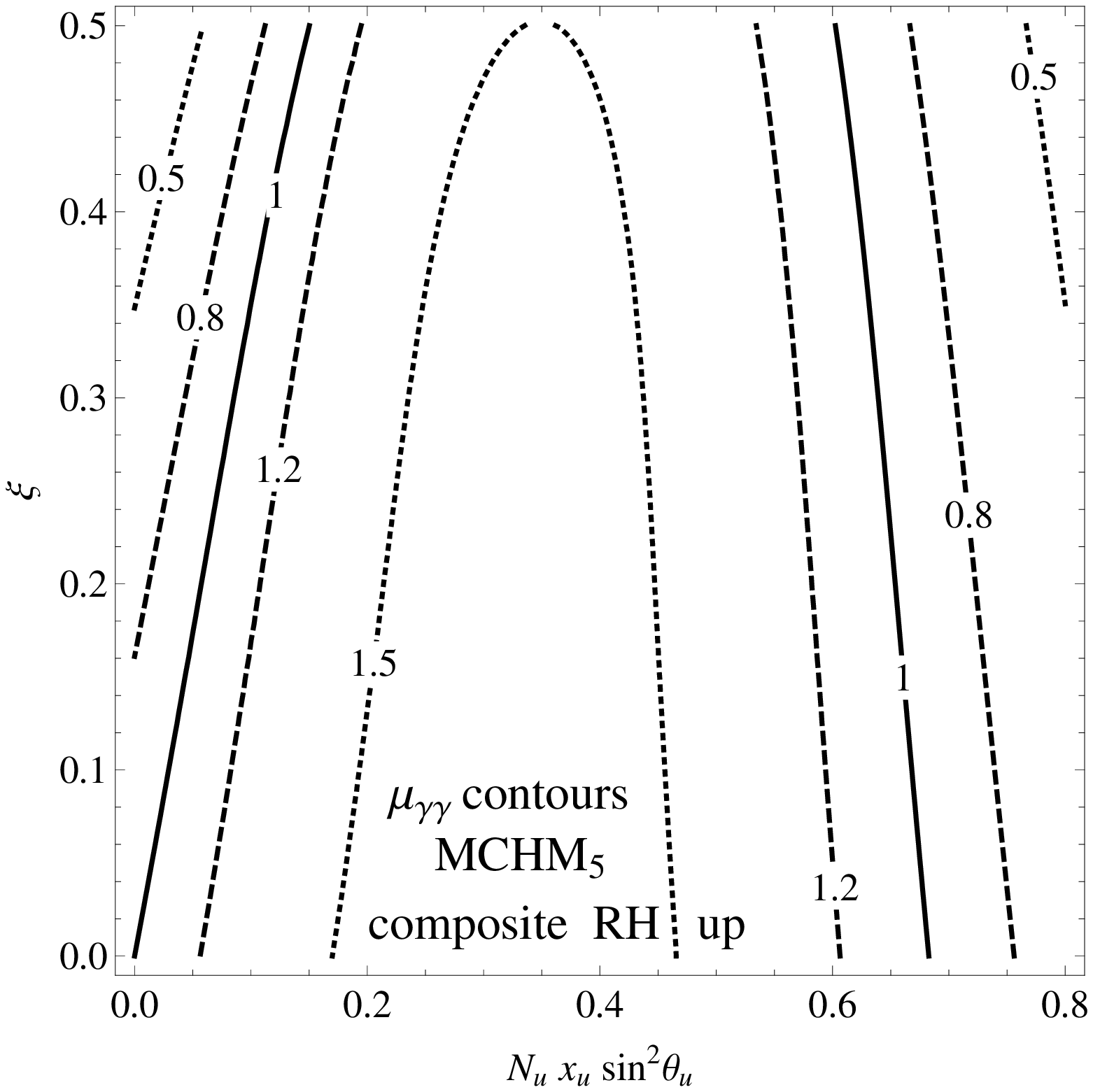}&\includegraphics[scale=0.5]{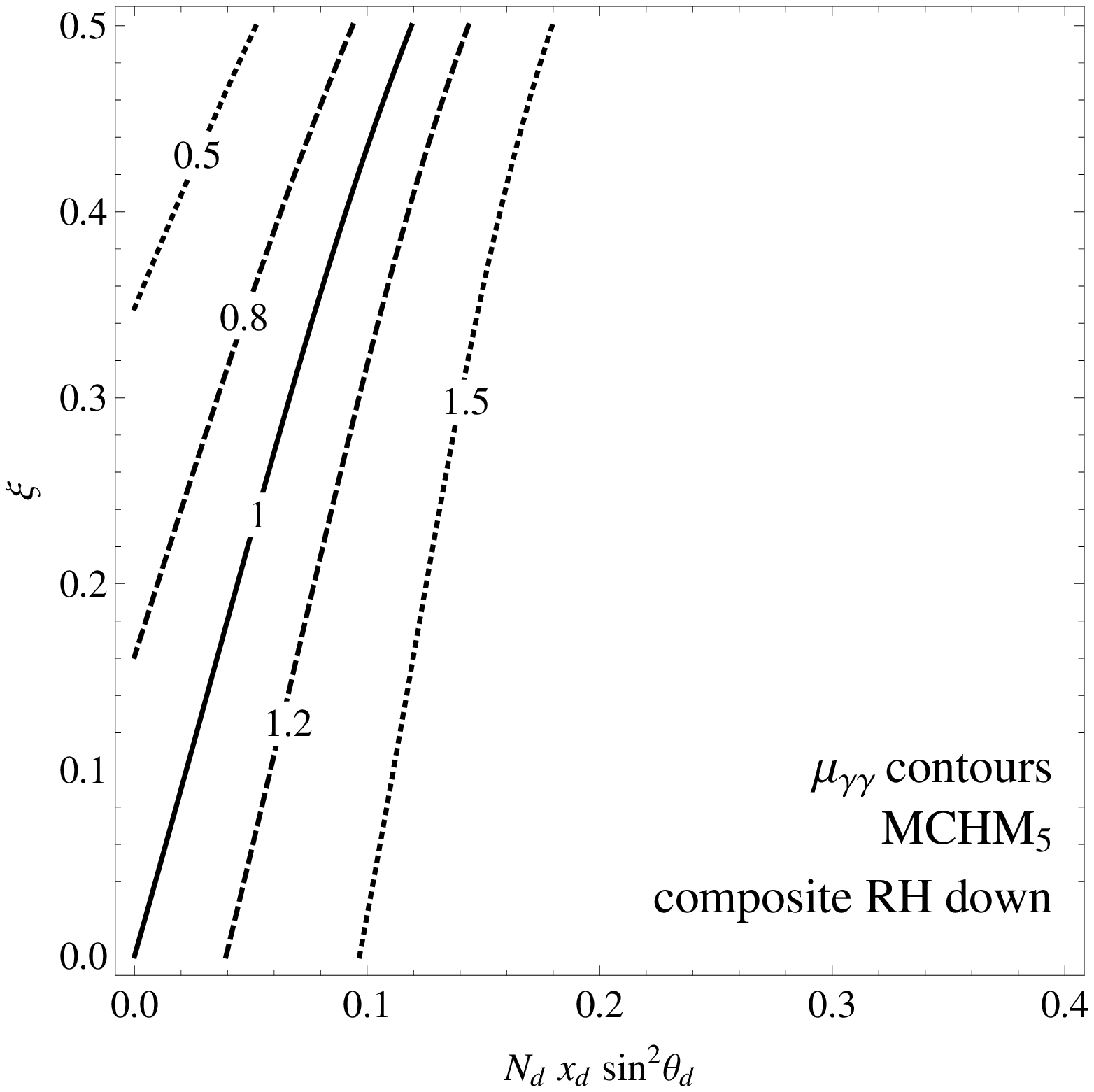}
\end{tabular}
\caption{Higgs signal strengths $\mu_{\gamma\gamma}$ in MCHM as a function of $\xi= v^2/f^2$ and $N_i x_i \sin^2\theta_i$ where $\sin\theta_i$  is the RH elementary/composite mixing and $x_i\equiv (Y_iv/M)^2$ and we set $r_i\equiv M/(Y_if)=1$. We considered two cases where either $i=u$ [left] or $i=d$ [right].}\label{mu5}
\end{figure}

\begin{figure}[t]
\centering
\begin{tabular}{cc}
\includegraphics[scale=0.5]{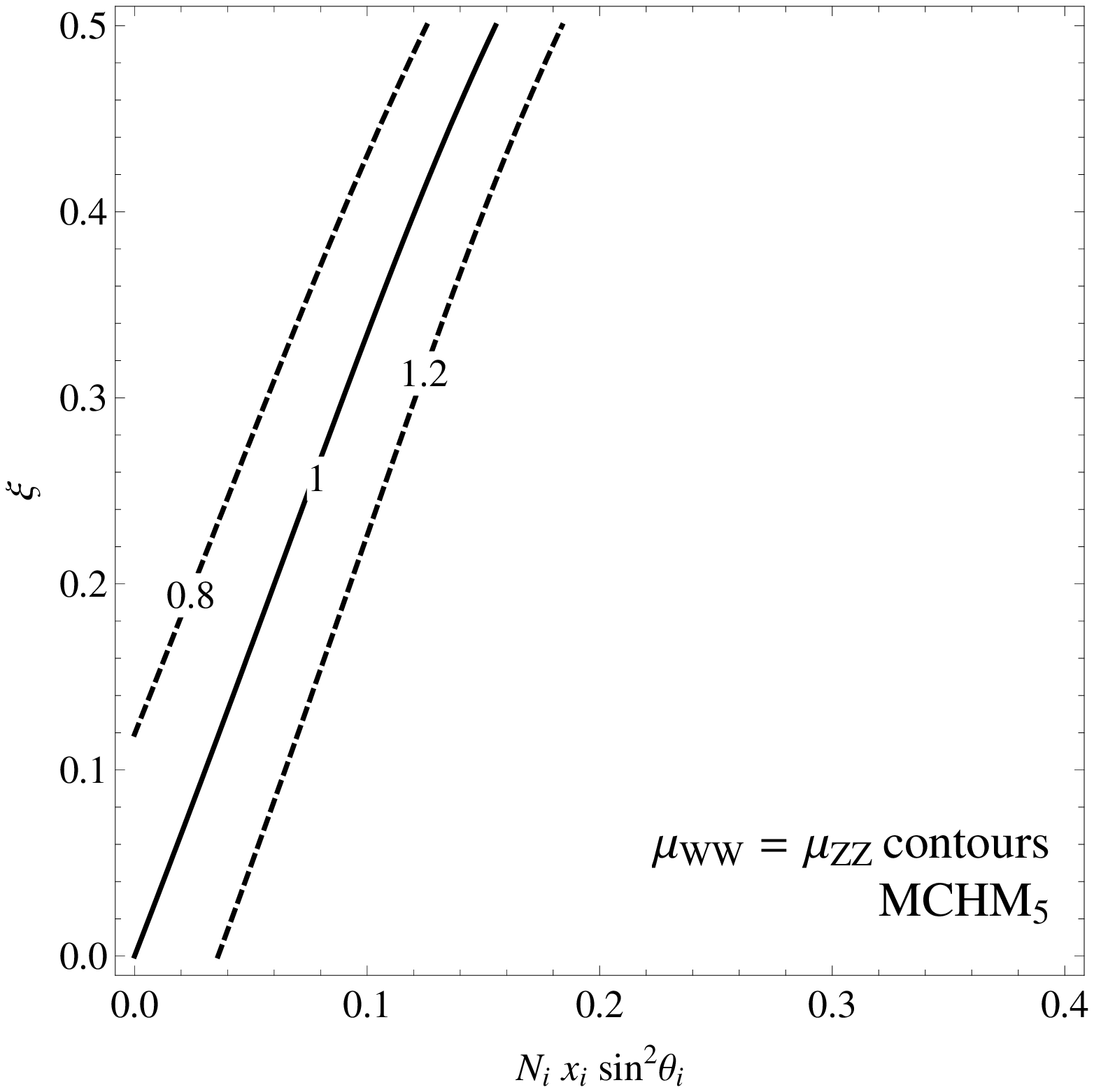}&\includegraphics[scale=0.5]{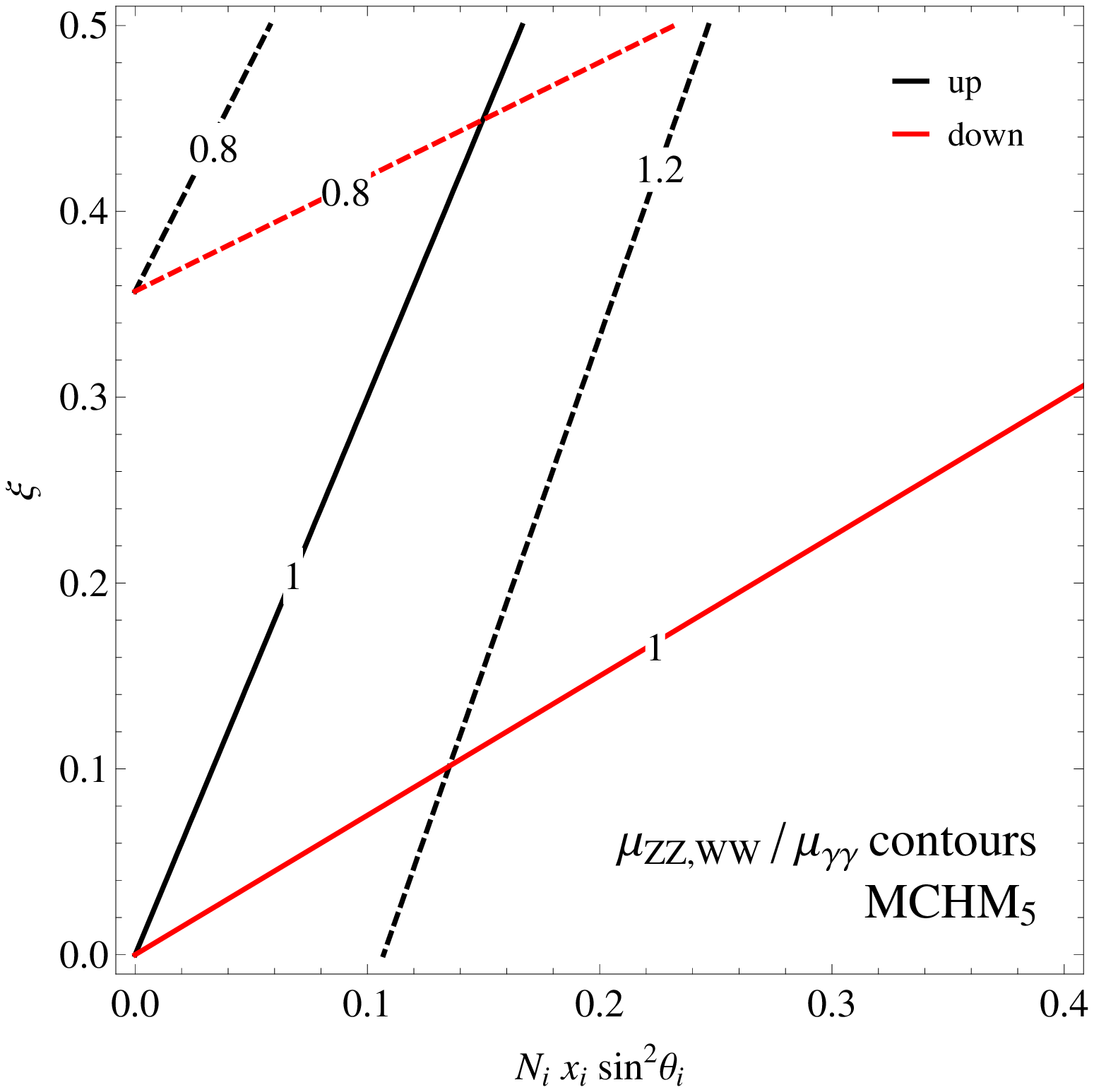}
\end{tabular}
\caption{Higgs signal strength $\mu_{ZZ,WW}$ [left] and $\mu_{ZZ,WW}/\mu_{\gamma\gamma}$ ratio [right] in MCHM as a function of $\xi= v^2/f^2$ and $N_i x_i \sin^2\theta_i$ ($i=u$ or $d$) where $\sin\theta_i$  is the RH elementary/composite mixing and $x_i\equiv (Y_iv/M)^2$ and we set $r_i\equiv M/(Y_if)=1$. We considered two cases where either $i=u$ (black contours) or $i=d$ (red contours).}\label{mu5zz}
\end{figure}

We show in Fig.~\ref{mu5split} the individual effect of Higgs non-linearities (left panel) and composite light flavors (right panel) on the $h\to \gamma\gamma$ and $h\to WW^*,ZZ^*$ signal strength.
We argued above that RH compositeness typically leads to an enhancement of the Higgs production cross-section, while, on the other hand, Higgs branching ratios in diphotons  tend to be suppressed. Thus, there is a region where the two effects compensate each other, leaving Higgs signal strengths close to their  standard predictions. We show on Fig.~\ref{mu5} the expected $\mu_{\gamma\gamma}$ in MCHM with $N_u$ (left panel) or $N_d$ (right panel) RH light flavors. Note that since down-type quarks contributions to $R_{\gamma\gamma}$ are suppressed by $Q_d^2/Q_u^2=1/4$ relative to up-type ones, the enhancement in gluon fusion is less compensated for relatively large RH down compositeness. The expected $\mu_{ZZ}=\mu_{WW}$ rate in MCHM  are shown on Fig.~\ref{mu5zz}. The latter are more sensitive to corrections in the production cross-section, as the $h\to ZZ^*,WW^*$ branching ratios are only mildly modified. 
%

\section{Conclusions} \label{sec:conclu}

We showed that, in composite pseudo Nambu--Goldstone boson (pNGB) Higgs models, flavor conserving Higgs observables at the LHC are rather sensitive to the degree of compositeness of the first two generation quarks, despite their {\it  a priori} negligible role in electroweak (EW) symmetry breaking. Large $\mathcal{O}(1)$ effects arise typically in models where the strong dynamics is not completely flavor anarchic but instead exhibits some flavor structures, since only the latter permits relatively composite right-handed (RH) light quarks. Therefore, flavor conserving Higgs physics can probe in a rather unique way the flavor structure of a broad class of composite Higgs models, at least at the qualitative level.

EW precision tests (EWPTs) at LEP and the recent Higgs rate measurements at the first LHC run did not find large deviations from New Physics (NP) beyond the Standard Model (SM). In the composite pNGB Higgs framework, the absence of NP evidence could be the result of either  a relatively high compositeness scale $f\gtrsim 800\,$GeV ({\it i.e} $\xi=v^2/f^2\lesssim 0.1$) or an accidental cancelation between large deviations from a lower compositeness scale and sizable contributions from TeV-scale resonances of the strong dynamics. 
The latter could arise for instance from spin one EW resonances. However their masses are typically constrained by the $S$ parameter to be above $\sim3\,$TeV, thus significantly restricting the size of spin one EW resonances effects on radiative Higgs couplings.
Lighter fermionic resonances, on the other hand, are allowed and could yield effects on Higgs couplings of the desired size. Although light partners for the composite top quark are expected from naturalness considerations, their effects on Higgs rates are rather model dependent and happen to be negligible in most minimal constructions due to a special structure of the fermion mass matrix. In contrast, sizable effects from TeV-scale composite partners inevitably arise provided (some of) the first two generation quarks are mostly composite fields.

Moreover, we find rather interesting that the most accurately measured Higgs rates could remain SM-like for moderate values of $\xi$ in the presence of a composite RH charm quark, without conflicting with EWPTs~\cite{FTRS,RediWeiler} or stringent flavor and dijet constraints~\cite{Exhil}. Future LHC measurements will directly probe the charm sector through charm-tagging based measurements. Any deviation from SM expectations in these searches would further shed light on the flavor structure of the strong dynamics at the TeV scale and thus potentially favor flavor ``order'' over complete anarchy. 

\section*{Acknowledgments}
We thank Kaustubh Agashe, Andrzej Buras, Roberto Contino, Maxim Perelstein, Stefan Pokorski, Francesco Riva, Heidi Rzehak, Raman Sundrum, Andrea Thamm and Robert Ziegler for discussions. 
C.G. is supported by the Spanish Ministry MICNN under contract FPA2010-17747 and by by the European Commission under the ERC Advanced Grant 226371 MassTeV and the contract PITN-GA-2009-237920 UNILHC.
GP is supported by grants from GIF, IRG, ISF and Minerva. 

\appendix

\section{SO(5)$/$SO(4) Essentials}\label{appSO5}

\subsection{``Pion'' Lagrangian}\label{Sigma}

We considered two-site models whose composite sector is a non-linear $\sigma$ model (nl$\sigma$m) with global SO(5)$\times$U(1)$_X$ symmetry. The non-linear $\Sigma$ field is 
\beq
\Sigma = \Sigma_0 \exp(-i\sqrt{2}h^{\hat a} T^{\hat a}/f)\,, \quad \Sigma_0 = (0,0,0,0,1)\,,	
\eeq
which  is subject to the non-linear constraint $\Sigma\Sigma^T=1$. $f$ is the SO(5) breaking scale, $T^{\hat a}$ are the 4 broken SO(5) generators (see below)
and $h_{\hat a}$ are 4 real NGBs. $\Sigma$ transform linearly as ${\bf 5}$ of  SO(5), while $h_{\hat a}$ transforms non-linearly under SO(5)$/$SO(4) but linearly as a ${\bf 4}$ of the unbroken SO(4) group. Upon mixing with the elementary sector, the Higgs radiatively develops a VEV breaking SO(4) to SO(3). By SO(4) rotation, one can align the Higgs component getting a VEV along the $\hat a =3$ direction: $h=h_3$. Hence, in  unitary gauge, {\it i.e.} removing the EW Goldstones, we have
\beq 
\Sigma = \Sigma_0\left(\begin{array}{ccccc}
1 &  &  &  &  \\
 & 1 &  &  & \\
 &  & \cos h/f &  & -\sin h/f \\
 &  &  & 1 & \\
 &  & \sin h/f &  & \cos h/f
\end{array}\right) = \left(0,0,\sin \frac{h}{f},0,\cos \frac{h}{f}\right)\,.
\eeq
The $\Sigma$ Lagrangian at two derivatives order is 
\beq 
\lag_{\rm kin} = \frac{f^2}{2}D_\mu \Sigma (D^\mu\Sigma)^\dagger \supset \frac{1}{2}\left(\partial_\mu h\right)^2-\frac{g^2f^2}{8}(\sin h/f)^2 W_\mu^2\, ,
\eeq
where $D_\mu$ is the SM covariant derivative, from which one finds 
\beq
c_H = -\frac{1}{2}c_r = \frac{1}{3f^2}\,.
\eeq

\subsection{Composite Fermion Representations}\label{Irreps}

\subsubsection{Vector representation}

A suitable basis for the 10 generators of SO(5) in the fundamental ${\bf 5}$ representation is
\beq
T^a_L &=& -\frac{i}{2}\left[\frac{\eps^{abc}}{2}\left(\delta^b_i\delta^c_j-\delta^b_j\delta^c_i\right)+\left(\delta^a_i\delta^4_j-\delta^a_j\delta^4_i\right)\right]\,,\label{TaL}\\
T^a_R &=& -\frac{i}{2}\left[\frac{\eps^{abc}}{2}\left(\delta^b_i\delta^c_j-\delta^b_j\delta^c_i\right)-\left(\delta^a_i\delta^4_j-\delta^a_j\delta^4_i\right)\right]\,,\label{TaR}\\
T^{\hat a} &=& -\frac{i}{\sqrt{2}}\left(\delta^{\hat a}_i\delta^5_j-\delta^{\hat a}_j\delta^5_i\right)\,,\label{Tcoset}
\eeq
where $T^a_{L,R}$ ($a=1,2,3$) generates the SU(2)$_{L,R}$ subgroups. 
Under the unbroken SO(4)$\sim$ SU(2)$_L\times$SU(2)$_R$  subgroup, the fundamental representation decomposes as ${\bf 5} = {\bf 1}+{\bf 4}$, with ${\bf 4} \sim ({\bf 2},{\bf 2})$. For $X=2/3$, we denote its  components as
\beq
{\bf 4}\sim ({\bf 2},{\bf 2}) = \left(\begin{array}{cc} D_\frac{1}{6}^+ & D_\frac{7}{6}^+\\
  D_\frac{1}{6}^- & D_\frac{7}{6}^-
\end{array}\right)\,,\quad {\bf 1}= S_\frac{2}{3}\,,
\eeq
where $D_\mathcal{Y}^\pm$ and $S_\mathcal{Y}$ denote, respectively, the $T_L^3=\pm 1/2$ components of a SU(2)$_L$ doublet and a SU(2)$_L$ singlet of hypercharge $\mathcal{Y}=T_R^3+X$.
The  embedding of $D_\frac{1}{6}$, $D_\frac{7}{6}$ and $S_\frac{2}{3}$ in an SO(5) vector follows from the definition of the generators in Eqs.~\eqref{TaL} and \eqref{TaR}
\beq
{\bf 5} = \frac{1}{\sqrt{2}}\left(D_\frac{1}{6}^--D_\frac{7}{6}^+,-i\left(D_\frac{1}{6}^-+D_\frac{7}{6}^+\right),D_\frac{1}{6}^++D_\frac{7}{6}^-,i\left(D_\frac{1}{6}^+-D_\frac{7}{6}^-\right),\sqrt{2}S_\frac{2}{3}\right)^T\,.
\eeq

\subsubsection{Adjoint representation}

The adjoint of SO(5) is a ${\bf 10}=({\bf 5}\times {\bf 5})_a$ which can be constructed out of the antisymmetric product of two fundamentals. The adjoint decomposes as ${\bf 10}={\bf 4} + {\bf 6}$ of SO(4), with ${\bf 6}\sim ({\bf 1},{\bf 3})+({\bf 3},{\bf 1})$. The components of the bidoublet and the triplets, respectively denoted as (assuming $X=2/3$)
\beq
({\bf 2},{\bf 2}) = \left(\begin{array}{cc} D_\frac{1}{6}^+ & D_\frac{7}{6}^+\\
  D_\frac{1}{6}^- & D_\frac{7}{6}^-
\end{array}\right)\,,\quad ({\bf 3},{\bf 1}) = \left(T_\frac{2}{3}^+,T_\frac{2}{3}^0,T_\frac{2}{3}^-\right)\,,\quad ({\bf 1},{\bf 3})= \left(S_\frac{5}{3},S_\frac{2}{3},S_{-\frac{1}{3}}\right)\,,
\eeq
where $T_\mathcal{Y}^{\pm,0}$ are the $T_L^3=\pm 1,0$ components of a SU(2)$_L$ triplet of hypercharge $\mathcal{Y}$, are embedded in the $5\times 5$ antisymmetric matrix as
\beq\label{10}
{\bf 10} = \frac{1}{2}\left(\begin{array}{cc} {\bf X} & {\bf D} \\ -{\bf D}^T & 0 \end{array}\right)\,,\quad {\rm where}\quad {\bf D} = \left(\begin{array}{c}D_\frac{1}{6}^--D_\frac{7}{6}^+ \\
 -i(D_\frac{1}{6}^-+D_\frac{7}{6}^+) \\
 D_\frac{1}{6}^++D_\frac{7}{6}^- \\
 i(D_\frac{1}{6}^+-D_\frac{7}{6}^-)\end{array}\right)\,,
\eeq
and ${\bf X}= {\bf X}_T+{\bf X}_S$, with
\beq
{\bf X}_T= \frac{1}{\sqrt{2}}\left(\begin{array}{cccc}
0 & i\sqrt{2}T_\frac{2}{3}^0& -(T_\frac{2}{3}^++T_\frac{2}{3}^-)& i(T_\frac{2}{3}^--T_\frac{2}{3}^+) \\
\cdot & 0 &i(T_\frac{2}{3}^--T_\frac{2}{3}^+)& T_\frac{2}{3}^-+T_\frac{2}{3}^+\\
\cdot  &\cdot   & 0 & i\sqrt{2}T_\frac{2}{3}^0 \\
\cdot & \cdot &\cdot & 0
\end{array}\right)\,,
\eeq
and
\beq
{\bf X}_S=\frac{1}{\sqrt{2}} \left(\begin{array}{cccc}
0 & i\sqrt{2}S_\frac{2}{3}& -(S_\frac{5}{3}+S_{-\frac{1}{3}})& i(S_\frac{5}{3}- S_{-\frac{1}{3}}) \\
\cdot & 0 &i(S_{-\frac{1}{3}}-S_\frac{5}{3})& -( S_\frac{5}{3}+S_{-\frac{1}{3}})\\
\cdot  &\cdot   & 0 & -i\sqrt{2}S_\frac{2}{3} \\
\cdot & \cdot &\cdot & 0
\end{array}\right)\,,
\eeq
where the $\cdot$ components are obtained from the antisymmetry property of ${\bf X}$.  

\subsubsection{Symmetric traceless matrix representation}

The $5\times 5$ symmetric traceless matrices form a ${\bf 14}$ representation of SO(5), whose SO(4) decomposition is ${\bf 14}={\bf 1}+{\bf 4}+{\bf 9}$, with ${\bf 9}\sim ({\bf 3},{\bf 3})$. The components of the singlet, bidoublet and bitriplet are respectively denoted as ${\bf 1}=S_\frac{2}{3}$,  
\beq
({\bf 2},{\bf 2}) = \left(\begin{array}{cc} D_\frac{1}{6}^+ & D_\frac{7}{6}^+\\
  D_\frac{1}{6}^- & D_\frac{7}{6}^-
\end{array}\right)\,,\quad {\rm and }\quad ({\bf 3},{\bf 3}) = \left(\begin{array}{ccc}
T_{-\frac{1}{3}}^+ & T_\frac{2}{3}^+ & T_\frac{5}{3}^+ \\
T_{-\frac{1}{3}}^0 & T_\frac{2}{3}^0  &T_\frac{5}{3}^0 \\
T_{-\frac{1}{3}}^- & T_\frac{2}{3}^- & T_\frac{5}{3}^-
\end{array}\right)\,,
\eeq
where $X=2/3$ was assumed. They are embedded in the $5\times 5$ symmetric traceless matrix as 
\beq
{\bf 14} = \frac{1}{2}\left(\begin{array}{cc} {\bf Y}& {\bf D} \\  {\bf D}^T& 0  \end{array}\right)+\frac{1}{2\sqrt{5}}{\rm diag}(1,1,1,1,-4)S_\frac{2}{3}\,,
\eeq
where ${\bf D}$ is given in Eq.~\eqref{10} and ${\bf Y}= {\bf Y}_{T_\frac{5}{3}}+{\bf Y}_{T_\frac{2}{3}}+{\bf Y}_{T_{-\frac{1}{3}}}$ with
\beq
{\bf Y}_{T_\frac{5}{3}}= \left(\begin{array}{cccc}
T_\frac{5}{3}^+ &  iT_\frac{5}{3}^+ &-T_\frac{5}{3}^0/\sqrt{2} & iT_\frac{5}{3}^0/\sqrt{2} \\
\cdot & -T_\frac{5}{3}^+ & -iT_\frac{5}{3}^0/\sqrt{2} & -T_\frac{5}{3}^0/\sqrt{2} \\
\cdot & \cdot & T_\frac{5}{3}^- & -iT_\frac{5}{3}^-\\
\cdot & \cdot & \cdot & -T_\frac{5}{3}^- 
\end{array}\right)\,,
\eeq
\beq
{\bf Y}_{T_\frac{2}{3}}= \left(\begin{array}{cccc}
-T_\frac{2}{3}^0 & 0  & \frac{1}{\sqrt{2}}(T_\frac{2}{3}^--T_\frac{2}{3}^+) & -\frac{i}{\sqrt{2}}(T_\frac{2}{3}^++T_\frac{2}{3}^-) \\
\cdot & -T_\frac{2}{3}^0& -\frac{i}{\sqrt{2}}(T_\frac{2}{3}^++T_\frac{2}{3}^-)&\frac{1}{\sqrt{2}}(T_\frac{2}{3}^+-T_\frac{2}{3}^-) \\
\cdot & \cdot &T_\frac{2}{3}^0 &0 \\
\cdot & \cdot & \cdot &T_\frac{2}{3}^0 
\end{array}\right)\,,
\eeq
and 
\beq
{\bf Y}_{T_{-\frac{1}{3}}}= \left(\begin{array}{cccc}
T_{-\frac{1}{3}}^- & -iT_{-\frac{1}{3}}^- & T_{-\frac{1}{3}}^0/\sqrt{2}& iT_{-\frac{1}{3}}^0/\sqrt{2}  \\
\cdot & -T_{-\frac{1}{3}}^-& -iT_{-\frac{1}{3}}^0/\sqrt{2}& T_{-\frac{1}{3}}^0/\sqrt{2}\\
\cdot & \cdot & T_{-\frac{1}{3}}^+& iT_{-\frac{1}{3}}^+\\
\cdot & \cdot & \cdot &  -T_{-\frac{1}{3}}^+
\end{array}\right)\,,
\eeq
where the $\cdot$ components are obtained from the symmetry property of ${\bf Y}$.

\section{EFT matching for higher fermionic representations}\label{1014matching}

We show here that similar effects as those presented in the main text are obtained in models where the fermionic resonances are embedded into larger SO(5) representations as the {\bf 10} or the {\bf 14}. For convenience, we report also here the results obtained for {\bf 5} representations.

Consider  the following strong sector Lagrangians for the resonances
\beq
\lag_{\rm strong}^{\bf 5} &= & \sum_{i=u,d} \bar\Psi^i(i\slashed D-M_i)\Psi^i -Y_i f (\bar{\Psi}_L^i \Sigma^T)(\Sigma\Psi_R^i)+\hc\,,\\
\lag_{\rm strong}^{\bf 10} &= & \bar\Psi(i\slashed D-M)\Psi-Y f (\Sigma \bar \Psi_L \Psi_R \Sigma^T)+\hc\,,\quad\\ 
\lag_{\rm strong}^{\bf 14} &=& \sum_{i=u,d} \bar\Psi^i(i\slashed D-M_i)\Psi^i-Y_i f (\Sigma\bar \Psi_L^i \Sigma^T)(\Sigma \Psi_R^i \Sigma^T)-Y_i' f (\Sigma\bar \Psi_L^i \Psi_R^i \Sigma^T)+\hc\, ,
\eeq
where $\Psi^{u,d}\sim{\bf 5}_{\frac{2}{3},-\frac{1}{3}}$, $\Psi\sim {\bf 10}_{\frac{2}{3}}$ and $\Psi^{u,d}\sim {\bf 14}_{\frac{2}{3},-\frac{1}{3}}$ of SO(5)$\times$U(1)$_X$, respectively. The case of the adjoint is treated slightly differently. In this case, only one multiplet of resonances is introduced (per generation) as the latter contains both the $S_{\frac{2}{3}}$ and $S_{-\frac{1}{3}}$ singlets required to induced up and down masses, and avoid at the same time large corrections at LEP for composite RH up and down quarks. As a consequence the elementary/composite mixings for the adjoint case are
\beq
-\lag_{\rm mix}\big|_{\bf 10} = \lambda_{q}\, \bar{q}_L D_{\frac{1}{6}} + \lambda_u\, \bar S_{\frac{2}{3}} {u}_R + \lambda_d \, \bar S_{-\frac{1}{3}} {d}_R +\hc\,,
\eeq
while, for the {\bf 5} and {\bf 14} cases, the mixings are given by Eq.~\eqref{Lmix}. For simplicity, we assume $M_u=M_d=M$ in the following and also set $Y_i'=0$ for the {\bf 14} case, as $Y_i$ alone is enough to reproduce the SM Yukawas.\\

Integrating out all the heavy fermionic states yields the following results for the Yukawas  for each case
\beq
y_u\big|_{\bf 5}=Y_u\sin\theta_q\cos\phi\sin\theta_u\,,\  y_u\big|_{\bf 10}=\frac{Y}{2\sqrt{2}}\sin\theta_q\sin\theta_u\,,\ 
y_u\big|_{\bf 14}=\frac{2\sqrt{2}Y_u}{\sqrt{5}}\sin\theta_q\cos\phi\sin\theta_u\,,
\eeq
where the sine of the LH mixing angles are 
\beq
\sin\theta_q&=&\left\{\begin{array}{ccc} 
\frac{\lambda_q}{\sqrt{\lambda_q^2+M^2}}&\,,&\Psi\sim{\bf 5}\,,{\bf 14}\,,\\
 \frac{\lambda_q}{\sqrt{\lambda_q^2+\left(M+\frac{Yf}{2}\right)^2}}&\,,&\Psi\sim{\bf 10}\,,
\end{array}\right.
\eeq
with $\lambda_q=\sqrt{\lambda_{q^u}^2+\lambda_{q^d}^2}$ and $\tan\phi = \lambda_{q^d}/\lambda_{q^u}$ whenever relevant, and the sine of the RH mixing angles are ($i=u,d$)
\beq
\sin\theta_i =\left\{\begin{array}{ccc}
\frac{\lambda_i}{\sqrt{\lambda_i^2+(M+Y_if)^2}}&\,,&\Psi\sim {\bf 5}\,,\\
\frac{\lambda_i}{\sqrt{\lambda_i^2+M^2}}&\,,&\Psi\sim {\bf 10}\,,\\
\frac{\lambda_i}{\sqrt{\lambda_i^2+\left(M+\frac{4Y_if}{5}\right)^2}}&\,,&\Psi\sim{\bf 14}\,.
\end{array}\right.
\eeq
Furthermore, the Wilson coefficients $c_y$ read $c_{y_i}=c_y^\Sigma + c_{y_i}^\Psi$ where the contribution from Higgs non-linearities are\footnote{In the SILH basis~\cite{SILH} where $c_r=0$, one finds $c_y^{\Sigma}\big|_{\bf 5}^{\rm SILH} = c_y^{\Sigma}\big|_{\bf 10}^{\rm SILH} =1/f^2$ and $c_y^{\Sigma}\big|_{\bf 14}^{\rm SILH} =7/(2f^2)$.}
\beq
c_y^\Sigma\big|_{\bf 5}=c_y^\Sigma\big|_{\bf 10} = \frac{4}{3f^2}\,,\quad c_y^\Sigma\big|_{\bf 14} = \frac{23}{6f^2}\,,
\eeq
and the composite resonance contributions are
\beq
c_{y_u}^\Psi\big|_{\bf 5} &=& \sin^2\theta_u\frac{Y_u\left(2M+Y_uf\right)}{fM^2}-\sin^2\theta_q\cos^2\phi\frac{Y_u\left(2M+Y_uf\right)}{2f\left(M+Y_uf\right)^2}\nonumber\\
&&-\sin^2\theta_u\sin^2\theta_q\cos^2\phi\frac{Y_u^2}{M^2}\left[1+\frac{M\left(2M+Y_uf\right)}{\left(M+Y_uf\right)^2}\right]\nonumber\\
&&+\sin^2\theta_u\sin^4\theta_q\cos^2\phi\frac{Y_u^2}{2M^2}+\sin^4\theta_u\sin^2\theta_q\cos^2\phi\frac{Y_u^2}{2\left(M+Y_uf\right)^2}\,,
\eeq
for the fundamental case,
\beq
c_{y_u}^\Psi\big|_{\bf 10}&=&-\sin^2\theta_u\frac{Y\left(4M+Yf\right)}{2f(2M+Yf)^2}+\sin^2\theta_q\frac{Y\left(4M+Yf\right)}{8fM^2}\nonumber\\
&& -\sin^2\theta_u\sin^2\theta_q\frac{Y^2}{8M^2}\left[1+\frac{2M\left(4M+Yf\right)}{\left(2M+Yf\right)^2}\right]\nonumber\\
&&+\sin^4\theta_u\sin^2\theta_q\frac{Y^2}{16M^2}+\sin^4\theta_q\sin^2\theta_u\frac{Y^2}{4\left(2M+Yf\right)^2}\label{cy10}\,,
\eeq
for the adjoint case, and
\beq
c_y^\Psi\big|_{\bf 14}&=&\sin^2\theta_u\frac{4Y_u(5M+2Y_uf)}{5fM^2}-
\sin^2\theta_q\cos^2\phi\frac{10Y_u\left(5M+2Y_uf\right)}{f\left(5M+4Y_uf\right)^2}\nonumber\\
&&-\sin^2\theta_u\sin^2\theta_q\cos^2\phi\frac{8Y_u^2}{5M^2}\left[1+\frac{10M\left(5M+2Y_uf\right)}{\left(5M+4Y_uf\right)^2}\right]\nonumber\\
&&+\sin^4\theta_u\sin^2\theta_q\cos^2\phi\frac{20Y_u^2}{\left(5M+4Y_uf\right)^2}+\sin^2\theta_u\sin^4\theta_q\cos^2\phi\frac{4Y_u^2}{5M^2}
\label{cy14}\,,
\eeq 
for the symmetric traceless case. The coefficients $y_d\big|_{\bf 5}$, $c_{y_d}^\Psi\big|_{\bf 5}$ and $y_d\big|_{\bf 14}$, $c_{y_d}^\Psi\big|_{\bf 14}$ are obtained from the coefficients $y_u\big|_{\bf 5}$, $c_{y_u}^\Psi\big|_{\bf 5}$  and $y_u\big|_{\bf 14}$, $c_{y_u}^\Psi\big|_{\bf 14}$, respectively,  with the replacements $Y_u,\sin\theta_u,\cos\phi\to Y_d,\sin\theta_d,\sin\phi$, and $y_d\big|_{\bf 10}$, $c_{y_d}^\Psi\big|_{\bf 10}$ are obtained from $y_u\big|_{\bf 10}$, $c_{y_d}^\Psi\big|_{\bf 10}$ with the replacements $\sin\theta_u,\sin\theta_q\to \sin\theta_d,-\sqrt{2}\sin\theta_q$.\\

After one-loop matching, we find that the relations 
\beq
c_g=\sum_{i=u,d}{\rm Re}[c_{y_i}^\Psi]\,,\quad
c_\gamma =\sum_{i=u,d}Q_i^2{\rm Re}[c_{y_i}^\Psi]\,, 
\eeq
still hold for all three cases. Note however that when $Y_i'\neq 0$ for the {\bf 14}  case the fermion mass determinant does not factorize as in Eq.~\eqref{detMfacto} and the above relations are not longer true.


\section{Loop Functions}\label{Loops}

We recall here the kinematical functions arising from the one-loop triangle diagrams of fermions ($A_{1/2}$) and charged gauge bosons ($A_1$) to the scalar to two gluons and/or photons amplitude~\cite{LET,spira}
\beq
A_{1/2}(\tau)=\frac{3}{2\tau}\left[1+(1-\tau^{-1})f(\tau)\right]\,,\quad
A_1(\tau)=\frac{1}{7\tau}\left[3+2\tau+3(2-\tau^{-1}) f(\tau) \right]\,,
\eeq
where $\tau\equiv m_h^2/(4m^2)$, $m$ being the loop particle mass, and 
\beq
f(\tau)=\left\{\begin{array}{cc} \left(\arcsin\sqrt{\tau}\right)^2 & (\tau\leq1)\,, \\
-\frac{1}{4}\left[\log\left(\frac{1+\sqrt{1-1/\tau}}{1-\sqrt{1-1/\tau}}\right)-i\pi\right]^2 & (\tau>1)\,.
\end{array}\right.
\eeq
For loop particles much heavier than the Higgs ($\tau\ll1$) the loop functions asymptote to unity as $A_{1/2}\simeq1+7\tau/30$ and $A_{1}\simeq1+22\tau/105$. Note that with this normalization of the loop function, the top and the $W$ contribute to the partial width of the Higgs into two photons proportionally to $Q_u^2 A_{1/2}(\tau_t) -7 A_{1}(\tau_W)/4$. For a discussion on how to include QCD and EW corrections, see Ref.~\cite{contino}.

\end{document}